\documentclass{elsart}
\usepackage{xspace}

\usepackage{natbib}

\usepackage{graphicx}

\usepackage{amssymb}

\newcommand{\MeV}{\ensuremath{\mbox{MeV}}\xspace}
\newcommand{\GeVc}{\ensuremath{\mbox{GeV}/c}\xspace}
\newcommand{\MeVc}{\ensuremath{\mbox{MeV}/c}\xspace}
\newcommand{\T}{\ensuremath{\mbox{T}}\xspace}

\newcommand{\cm}{\ensuremath{\mbox{cm}}\xspace}
\newcommand{\mm}{\ensuremath{\mbox{mm}}\xspace}

\newcommand{\ns}{\ensuremath{\mbox{ns}}\xspace}
\newcommand{\m}{\ensuremath{\mbox{m}}\xspace}

\newcommand{\ms}{\ensuremath{\mbox{ms}}\xspace}
\newcommand{\micros}{\ensuremath{\mu \mbox{s}}\xspace}

\newcommand{\dedx}{\ensuremath{\mbox{d}E/\mbox{d}x}\xspace}
\newcommand{\pip}{\ensuremath{\pi^+}\xspace}
\newcommand{\pim}{\ensuremath{\pi^-}\xspace}

\newcommand{\dzero}{\ensuremath{d_0}\xspace}
\newcommand{\dzeroprime}{\ensuremath{d'_0}\xspace}

\newcommand{\evtspill}{\ensuremath{N_{\mathrm{evt}}}\xspace}
\newcommand{\pt}{\ensuremath{p_{\mathrm{T}}}\xspace}
\newcommand{\tht}{\ensuremath{\theta}\xspace}

\def\be{\begin{equation}}
\def\ee{\end{equation}}
\def\bea{\begin{eqnarray}}
\def\eea{\end{eqnarray}}

\begin{document}

\begin{frontmatter}



\title{Dynamic Distortions in the HARP TPC: observations, measurements, modelling and corrections}


\author[in1]{A. Bagulya}
\author[in2]{A. Blondel}
\author[in2,in3]{S. Borghi}
\author[in4]{G. Catanesi}
\author[in5]{P. Chimenti}
\author[in6]{U. Gastaldi}
\author[in7,cor1]{S. Giani}
\author[in8]{V. Grichine}
\author[in7,in9]{V. Ivanchenko}
\author[in10]{D.~Kolev}
\author[in7]{J. Panman}
\author[in4]{E. Radicioni}
\author[in10]{R.~Tsenov}
\author[in11]{I.~Tsukerman}
\corauth[cor1]{Corresponding author e-mail: Simone.Giani@cern.ch}

\address[in1]{Institute for Nuclear Research, Moscow, Russia}
\address[in2]{ Universit\'{e} de Gen\`{e}ve, Switzerland}
\address[in3]{ Current address University of Glasgow, UK}
\address[in4]{INFN, Bari, Italy}
\address[in5]{INFN, Trieste, Italy}
\address[in6]{Laboratori Nazionali di Legnaro dell' INFN, Legnaro, Italy}
\address[in7]{CERN, Geneva, Switzerland}
\address[in8]{P. N. Lebedev Institute of Physics (FIAN),
Russian Academy of Sciences, Moscow, Russia}
\address[in9]{On leave of absence from Ecoanalitica, Moscow State University,
  Moscow, Russia}
\address[in10]{Faculty of Physics, St. Kliment Ohridski University, Sofia, Bulgaria}
\address[in11]{ITEP, Moscow, Russian Federation}

\begin{abstract}

The HARP experiment was designed to study hadron production in
proton--nucleus collisions in the energy range of
1.5~\GeVc--15~\GeVc. The experiment was made of two spectrometers, a
forward dipole spectrometer and a large-angle solenoid spectrometer. In
the large-angle spectrometer the main tracking and particle
identification is performed by a cylindrical Time Projection Chamber
(TPC)  which suffered a number of shortcomings later addressed in the
analysis.  In this paper we discuss the effects of time-dependent
(\emph{dynamic})  distortions of the position measurements in the TPC
which are due to a build-up of ion charges in the chamber during the
accelerator spill.  These phenomena have been studied both theoretically
and experimentally,  and a correction procedure has been developed.
First, the dynamics of the positive ion cloud and of the full
electrostatics of the field-cage system have been modelled with a
phenomenological approach  and a general correction procedure has been
developed and applied to all data settings.  Then,  the correction
procedure has been benchmarked experimentally by  means of recoil
protons in elastic scattering reactions, where the track coordinates are
precisely predictable from simple kinematical considerations.  After
application of the corrections for dynamic distortions the corrected
data have a performance equal to data where the dynamic distortions are
absent.  We describe the theoretical model, the comparison with the
measurements, the distortion correction method and the results obtained
with experimental data.

\end{abstract}




\end{frontmatter}

\section{Introduction}
\label{sec:intro-harp}

The HARP experiment \cite{ref:harp:spsc99,ref:harp:detector}
was designed to study hadron production in proton--nucleus
collisions in the energy range of 1.5~\GeVc--15~\GeVc. 
The main aim of the experiment 
is to provide pion production data for the calculation of neutrino
fluxes in conventional neutrino beams at accelerators, to provide data for
extended air shower simulations and for prediction of the atmospheric neutrino
flux, as well as to provide input to the quantitative design of a future
neutrino factory.  

The experiment was made of two spectrometers:
\begin{itemize}
\item  A forward dipole spectrometer with
planar drift chambers for the particle tracking and a time-of-flight (TOF) scintillator 
wall, a
Cherenkov detector and an electromagnetic calorimeter for particle
identification (PID).
\item  A large-angle solenoid spectrometer where the main
tracking and PID is performed by a cylindrical Time Projection Chamber (TPC) occupying most
of the radial space of the solenoid magnet. 
The TPC provides track, momentum and vertex measurements for all outgoing charged
particles in the angular range from 20$^\circ$ to 135$^\circ$ with respect to
the beam axis. In addition, it provides particle identification by recording the
particle's energy loss  in the gas (\dedx). The PID
capabilities of the TPC detector are complemented by a set of multi-gap 
RPCs (Resistive Plate Chambers)
serving as TOF detectors and surrounding the TPC.  

\end{itemize}

Data analysed with the large-angle spectrometer have been published in 
Refs.~\cite{ref:tantalum,ref:harp:art1,ref:harp:art2,ref:harp:art3}

\section{The HARP TPC}
\label{sec:intro-tpc}

\begin{figure}[tbp]
\vspace{9pt}
\begin{center}
\includegraphics[width=0.65\textwidth]{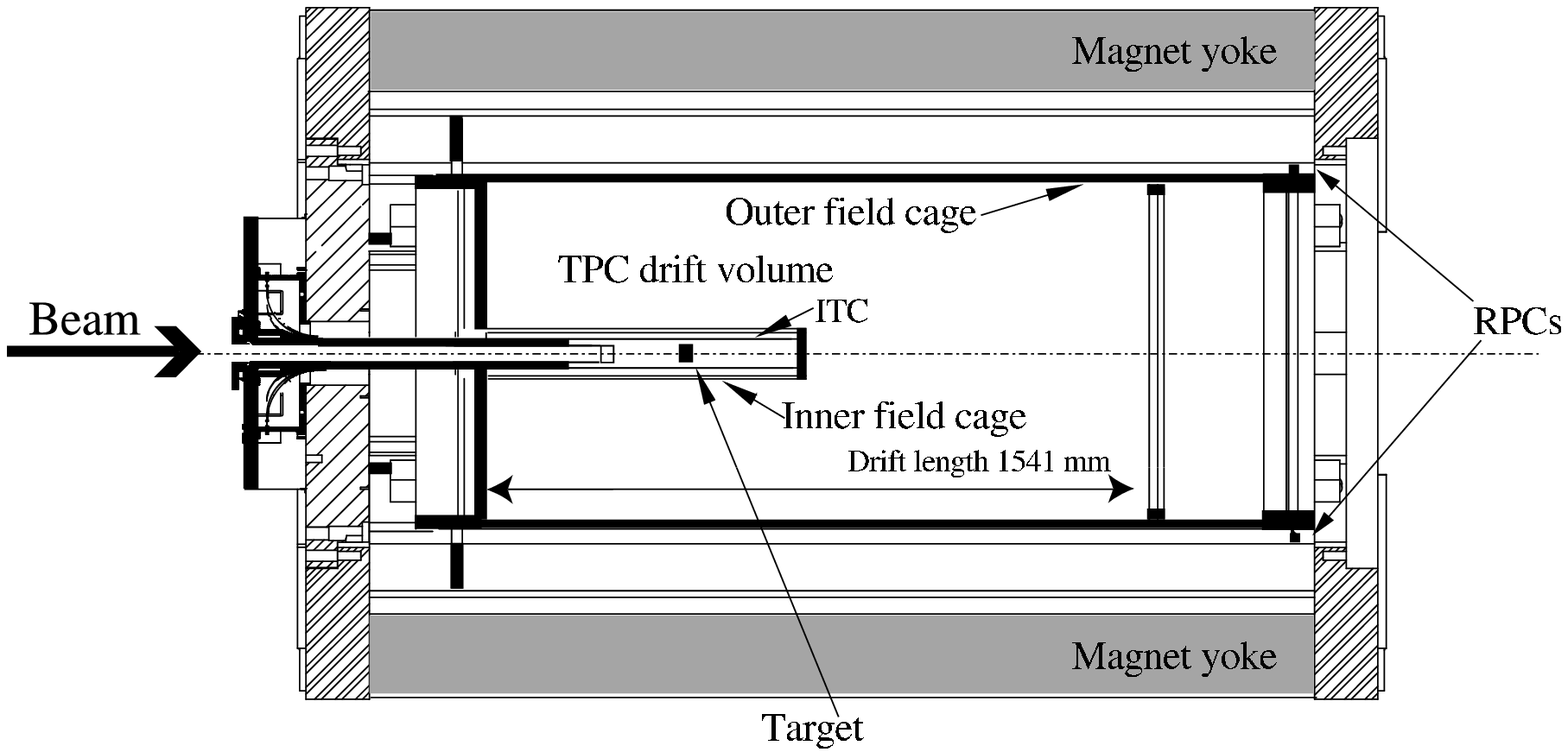}
\includegraphics[width=0.3\textwidth]{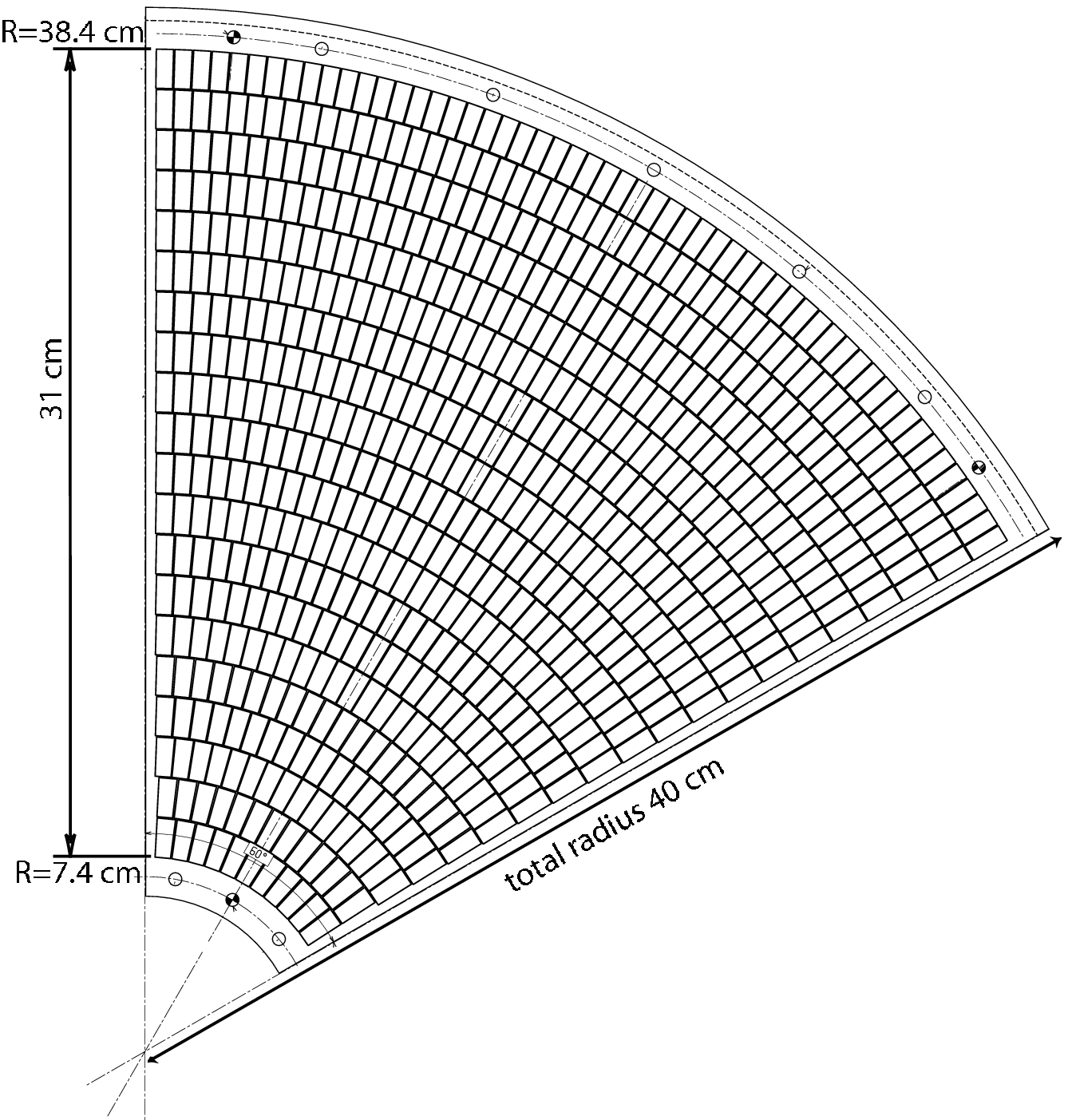}
\end{center}
\caption{ Schematic layout of the  TPC. The beam enters from the left. Starting
from the outside, first the return yoke of the magnet is seen, closed with an
end-cap at the upstream end, and open at the downstream end.  
The inner field cage is visible as a short cylinder entering
from the left. The ITC trigger counter and the target holder are inserted in the
inner field cage.  The RPCs (not drawn) are positioned between the outer field
cage and the coil.  The drift length is delimited on the right by the
 cathode plane and on the left by the anode wire planes.
On the right the mechanical drawing of a sector of
 the TPC, the layout of the pads is indicated. 
} 
\label{fig:tpc}
\end{figure}

The schematic layout of the HARP TPC is shown in Fig.~\ref{fig:tpc}.
The TPC is positioned inside the solenoid magnet, providing  a
magnetic volume with a diameter of 0.9~\m, a length of 2.25~\m  and a field of
0.7~\T in the main sensitive volume. The magnet was previously used for the 
R\&D of the ALEPH experiment's TPC \cite{ref:tpc90} 
and later modified for HARP. 
The downstream end of the return yoke was left open to minimize materials 
encountered by secondary particles emerging from the TPC in the direction of
 the forward spectrometer. At the upstream end there is a small cylindrical 
hole in the end-cap yoke for
the passage of the incident beam and to leave space to insert the inner
trigger cylinder (ITC)
and target holder inside the inner field cage (IFC). 
The drift volume is 1541~\mm long with a nominal electric field gradient of 111~V/\cm.
Given the drift velocity of the chosen gas mixture under these operating
conditions, the maximum drift time is approximately 30~\micros. The induced
charge from the gas amplification at the anode wires is measured using a pad
plane, subdivided in six sectors; the anode wires are strung onto the six spokes
defining the sectors. The pads are organized in 20 concentric rows, each pad
being connected to an individual pre-amplifier. The pad dimensions are 
6.5~\mm~$\times$~15~\mm and the number of pad ranges from 11 per row per sector at
the inner radius to 55 at the outer radius. 
The pad-charges are sampled into charge time-series by one 
Flash-ADC (FADC) per pad, with a sampling interval of 100~\ns. 
The total Data Acquisition (DAQ) readout time is
500~\micros to 1000~\micros per event depending on the event size.

During the analysis, after unpacking the FADC values, time-series are
organized in $R\phi$ clusters.
The clusters are assigned to tracks by a tree-based algorithm for pattern recognition
which used a general three-dimensional binary search method for fast look-up of clusters
~\cite{ref:uiterwijk}. 

Once clusters are assigned to a track, a helix fit is performed. The
fitting procedure is based on the algorithm developed by the ALEPH
Collaboration~\cite{ref:aleph} with slight
modifications~\cite{ref:morone}, 
{\em e.g.} the possibility to fit tracks which spiral for more than
2$\pi$~\cite{ref:silvia:thesis}. 
The fit consists of two consecutive steps: a circle-fit  in the $x$--$y$
plane\footnote{The Cartesian coordinates $x$ and $y$ are the coordinates perpendicular
to the nominal magnetic field.},
based on a least-square method~\cite{ref:chernov},
and a subsequent straight line fit in the $z$--$s_{xy}$ plane\footnote{The $s_{xy}$
coordinate is defined as the arc length along the circle in the $x$-$y$ plane
between a point and the impact point.}. 
The two fitting steps allow the five parameters which uniquely define
the helix to be determined.  
The code uses the same naming and
sign conventions as in the TASSO and ALEPH software~\cite{ref:aleph} with a
particle direction associated to the motion along the helix itself.

The analysis revealed that the TPC suffered from a number of operational
problems which were discovered, one after the other, during and after
the data taking: 

\begin{enumerate}
\item large excursions of the gains of the pad pre-amplifiers;
\item a relatively large number of dead or noisy pads;
\item large pad gain variations with time;
\item static distortions caused by the inhomogeneity of the magnetic
  field, an accidental HV mismatch between the inner and outer field
  cage;
\item cross-talk between pads caused by capacitive coupling between
  signal lines in the multilayer printed boards; 
\item dynamic distortions caused by build-up of ion-charge density in 
  the drift volume during the 400~ms long beam spill.
\end{enumerate}

The corrections for the first four effects are described in
Refs.~\cite{ref:harp:detector,ref:tantalum,ref:silvia:thesis}. A detailed
discussion of
the cross-talk effect can be found in Ref.~\cite{ref:lara}.  
The dynamic distortion effect and its corrections are presented in this paper.

\section{Evidence for TPC \emph{Dynamic Distortions} and overall
characteristics}
\label{sec:evidence}

\begin{figure}
\begin{center}
\includegraphics[width=0.45\textwidth]{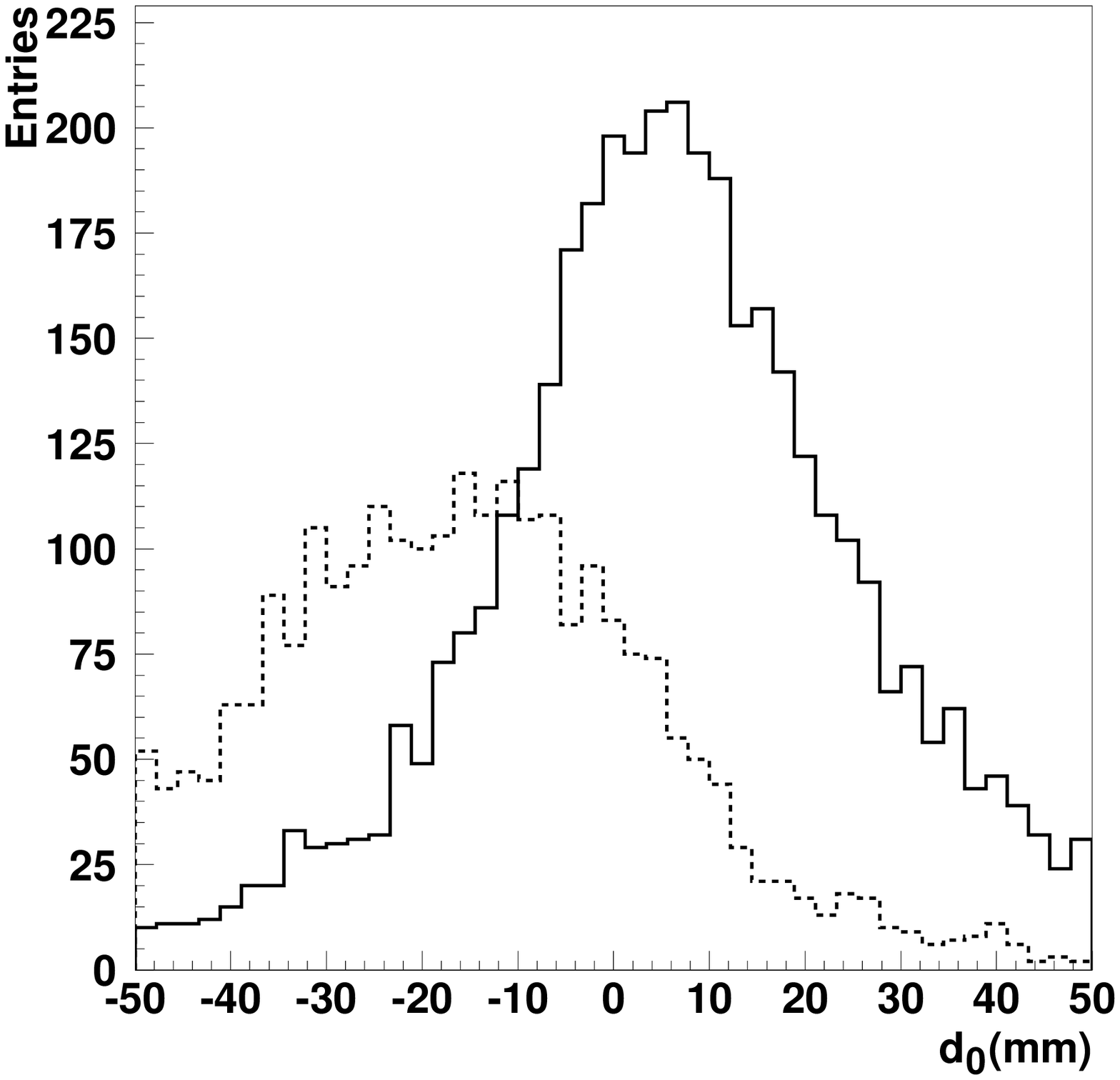}
\includegraphics[width=0.45\textwidth]{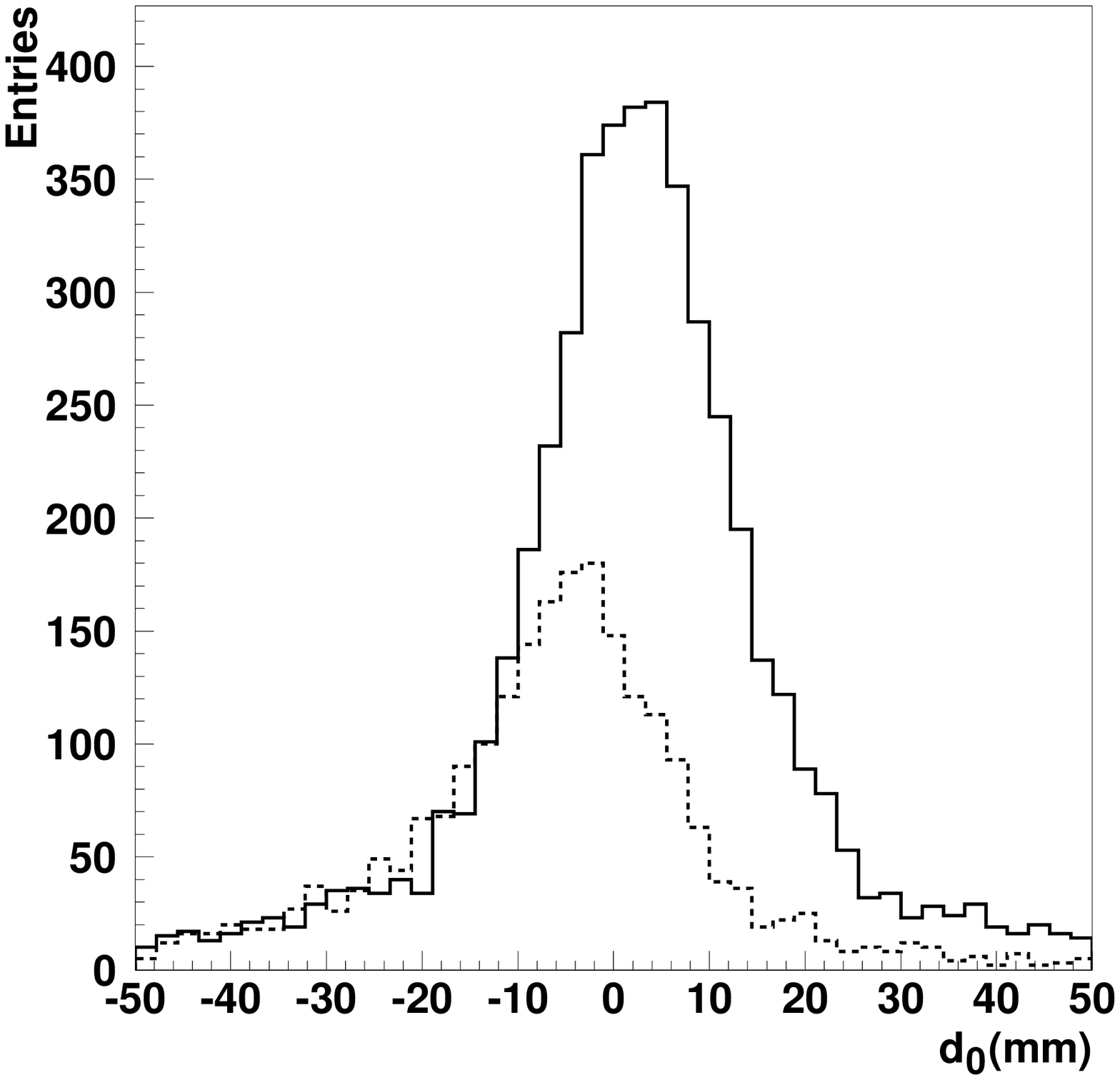}
\end{center}
\caption{The $d_0$ distributions in a 8~\GeVc beam exposure on a
 5\%~$\lambda_{\mathrm{I}}$ Be target. Left panel: run 9540. Right panel:
 run 9455.
The histograms drawn with continuous lines are for positive particles
 and the the ones with dotted lines for negative particles.
The two runs are taken close to each
other in time, one just before and the other after re-tuning of the beam.
}
\label{fig:distortion:evidence}
\end{figure}

In a situation where an incoming beam hits a target at the centre of
a rotationally symmetric TPC one expects the tracks of produced
particles to a good approximation to emerge from the centre of the TPC. 
We then define \dzero to be the impact point of the tracks in the $xy$
plane, i.e. the minimum distance between the track and the $z$-beam axis
in the $xy$ plane. By convention, its sign indicates whether the helix
encircles the $z$-beam axis (positive sign) or not (negative sign).
The distribution of \dzero is then expected to be symmetric around the
origin. 
The presence of distortions in the TPC can modify this distribution.
Thus, the distribution of the distance of closest approach of the helix
to the nominal axis of the TPC is a measure of distortions in the TPC. 
The \dzero distribution of TPC tracks was found to show a difference
between the peaks for tracks with opposite curvature, wih a separation
depending on the beam tuning and intensity~\cite{ref:silvia:thesis}. 
This effect is shown in Fig.~\ref{fig:distortion:evidence}. 
The two panels of the figure show two different runs of the same setting
(8~\GeVc on a 5\%~$\lambda_{\mathrm{I}}$ Be target) taken close to each
other in time, one just before and the other after re-tuning of the
beam.

The presence of the effect persisted even after correcting for static
distortion due to voltage misalignment between the inner and outer field
cages, the effect of which was shown to be much less important.

A better measure of the impact parameter, namely the track impact
distance with respect to the trajectory of the incoming beam particle,  
\dzeroprime\footnote{The \dzeroprime sign indicates
if the helix encircles the beam particle trajectory (positive sign) or not (negative
sign)}, was found to be a very sensitive probe to measure the
distortion strength.
The difference between \dzeroprime and \dzero may be large due to the
relatively large width of the beam spot at the target ($\approx$5~\mm)
and due to the fact that the beam was not always centred.
  
\begin{figure}
\begin{center}
\includegraphics[width=0.49\linewidth]{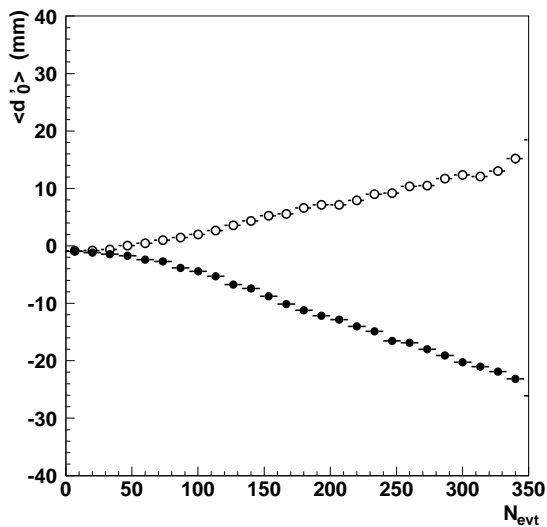}
\end{center}
\caption{Example of monitoring of the dependence of the distortion on the time during the spill. \dzeroprime as function of the event number in the spill for a 3 \GeVc beam exposure on a Ta target.
Open circles: positively charged particles; closed circles: negatively
 charged particles.}
\label{fig:distortion:spilldependence}
\end{figure}

The influence of the distortions can be monitored using its average
value $\langle \dzeroprime \rangle$ as shown in Ref.~\cite{ref:silvia:thesis}. 
In Fig.~\ref{fig:distortion:spilldependence} this quantity is displayed
separately for positively and negatively charged pion tracks 
as a function of the event number within the spill \evtspill.  
A similar benchmark was used in Ref.~\cite{ref:star}
for distortions observed in the STAR TPC.
Due to the sign-convention, the dynamic distortions shift the \dzeroprime 
value for particle tracks of positive and negative charge in opposite direction.
The dependence of \dzeroprime shows that a distortion effect builds up during the
spill, testifying its \emph{dynamic} character.
The absence of distortions at the beginning of the spill, and the
increasingly larger distortions with opposite sign for oppositely
charged particles during the spill, explains the peak structure
initially observed in the \dzero distribution. 
The curvature of high momentum tracks changes sign and therefore
their \dzeroprime migrates to large values with opposite sign.

The  evolution of the \dzeroprime distortion during the spill suggests
a slowly drifting cloud of positive ions to be the cause of the track
distortions.
In addition, the absence of a saturation plateau suggests
that the ion build-up does not reach its maximum before the end of the
spill.
It is therefore expected that a cloud of positive ions generated around
the wires of the anode grid of the TPC grows progressively with time
during the spill. The ion cloud would have approximately a torical shape
with inner and outer diameter limited by the inner and outer field cages
of the TPC. The front of the ion cloud does not reach the TPC  HV
cathode before the end of the spill. At the end of the beam spill the
production of positive ions stops. During the inter-spill time the full
cloud drifts to the HV cathode and vanishes completely before the start
of the following spill.  The minimum time between spills is ensured
to be 2 s by the operation of the PS.

It was observed that the overall effect of the dynamic distortion in
$R\phi$ was opposite in sign to the one induced by the  
static distortions due to the aforementioned HV mismatch.
Via a straightforward $E \times B$ calculation it is possible to
conclude that the radial component of the electric field due to dynamic
distortions has to be directed predominantly outwards.

The hypothesis that dynamic distortions are caused by the build-up of positive ions in
the drift volume during the 400~\ms long beam spill makes it easier to understand why
changes in the beam parameters (intensity, steering, focus) cause an increase or
decrease in the dynamic distortions: the large amount of material around the
target is likely to produce many low-energy secondary particles very close to the 
inner field cage whenever it is hit by a sizable beam halo. 
After the end of the beam spill, the initial conditions are
re-established by the fact that the inter-spill time is large
enough to drain all ions to the HV cathode.

Although charge build-up is a common phenomenon in TPCs, in the case of HARP the
large amount of material before and around the target -- very close to the beam
axis -- and the way the beam has often been tuned (too intense and/or not enough
collimated) in some of the setting makes its occurrence rather difficult to deal
with. In addition, the strongly inhomogeneous distortion of the 
electric field (larger at smaller radius) complicates the dynamic modelling of the
actual field lines and of the ionization charge trajectories.

\section{Benchmarking the distortion effects using elastic scattering}
\label{sec:elastics-effect}

Elastic scattering interactions of protons and pions on hydrogen provide events
where the kinematics are fully determined by the scattering angle of the forward
scattered beam particle. These kinematic properties were exploited to provide a 
known \emph{beam} of protons pointing into
the TPC sensitive volume. 
Data sets taken with liquid hydrogen targets at beam momenta from
3~\GeVc to 8~\GeVc were used for this analysis. 

A good fraction of forward elastically-scattered protons or pions 
enter into the acceptance of the forward spectrometer, where the full
kinematics of the event can be constrained. In particular, the direction
and momentum of the recoil proton can be precisely predicted. Selecting events
with
one and only one track in the forward direction and requiring that the measured
momentum and angle are consistent with an elastic reaction already provides an
enriched sample of elastic events. By requiring that only one barrel RPC hit is
recorded at the position predicted for an elastic event (the precision of the 
prediction from the forward spectrometer is within the RPC pad size) and within
a time window consistent with a proton time-of-flight, we obtained a $\simeq$99\% pure
sample of recoil protons in the TPC volume and with known momentum vector.

At beam momenta in the range 3~\GeVc--8~\GeVc the protons which are
tagged by accepted forward beam particles point into the TPC with angles
of $\approx 70^{\circ}$ with respect to the beam direction. 
The beam counters provide a direct measurement of the incoming particle
direction and of the scattering vertex coordinates in the target transverse plane.
Once a clean sample of elastic-scattering events is isolated, by  using detectors 
that are all independent from the TPC, the absolute efficiency of the  track 
finding and fitting procedure can be measured  (this allows the Monte Carlo 
calculations of the TPC detection efficiency to be benchmarked),  and  both 
the direction of emission and the momentum at the scattering vertex of the 
proton which traverses the TPC at large angle are determined for each event. 
After correction for energy loss and multiple scattering the complete track 
trajectory is determined. It  can be used both for comparison with the track 
points of the proton track reconstructed from the TPC data not corrected for 
distortions, to measure directly the effect of dynamic distortions and for 
benchmarking the proton track reconstructed from TPC data after corrections
 of dynamic distortions \cite{ref:INFN-LNL}.
This fact has been used for a direct measurement of the distortions, as it 
will be described in Sec.~\ref{sec:elastics-shift}. 

\begin{figure}[tbp]
  \begin{center}
    \includegraphics[width=0.45\linewidth]{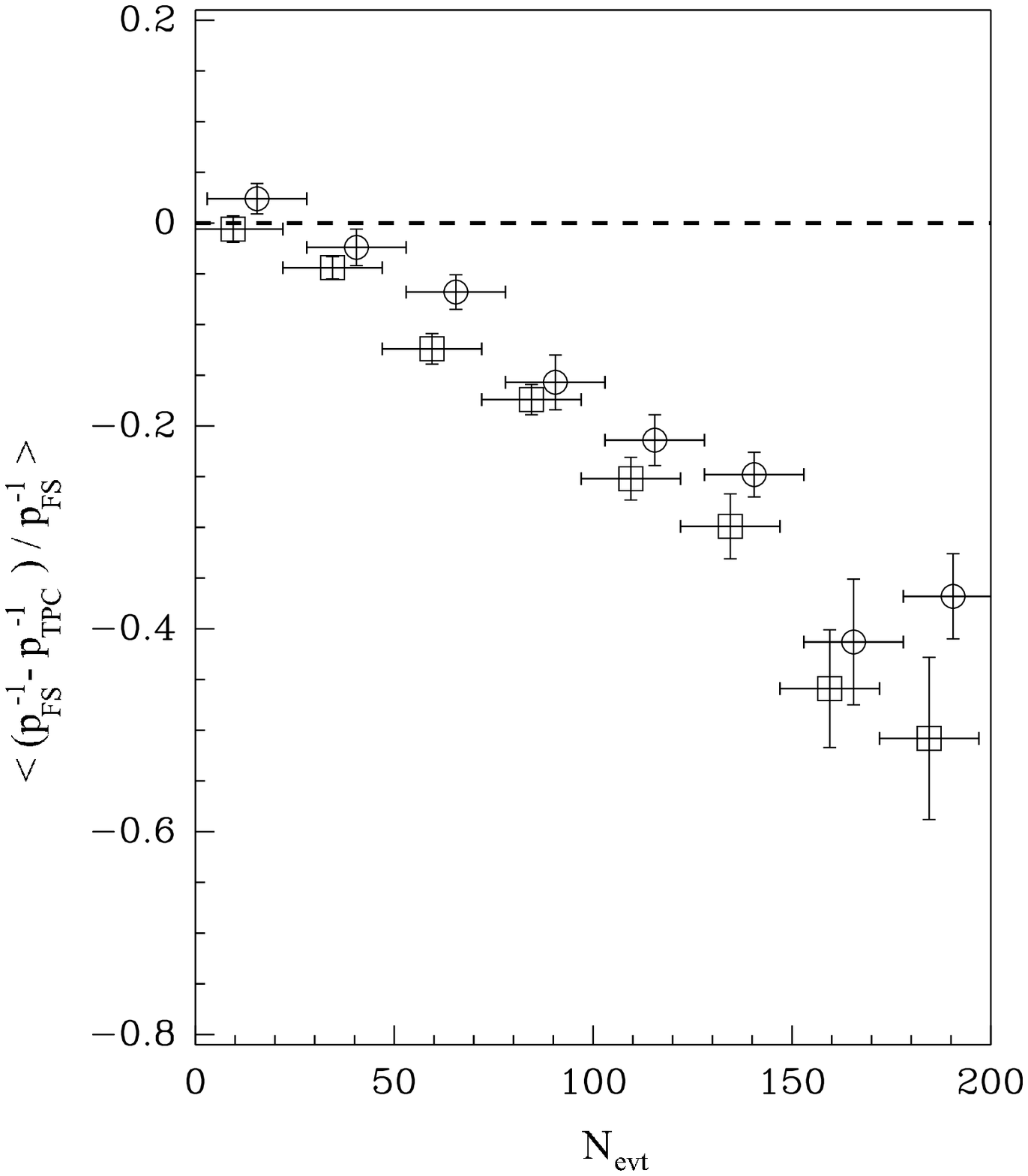}
    \includegraphics[width=0.45\linewidth]{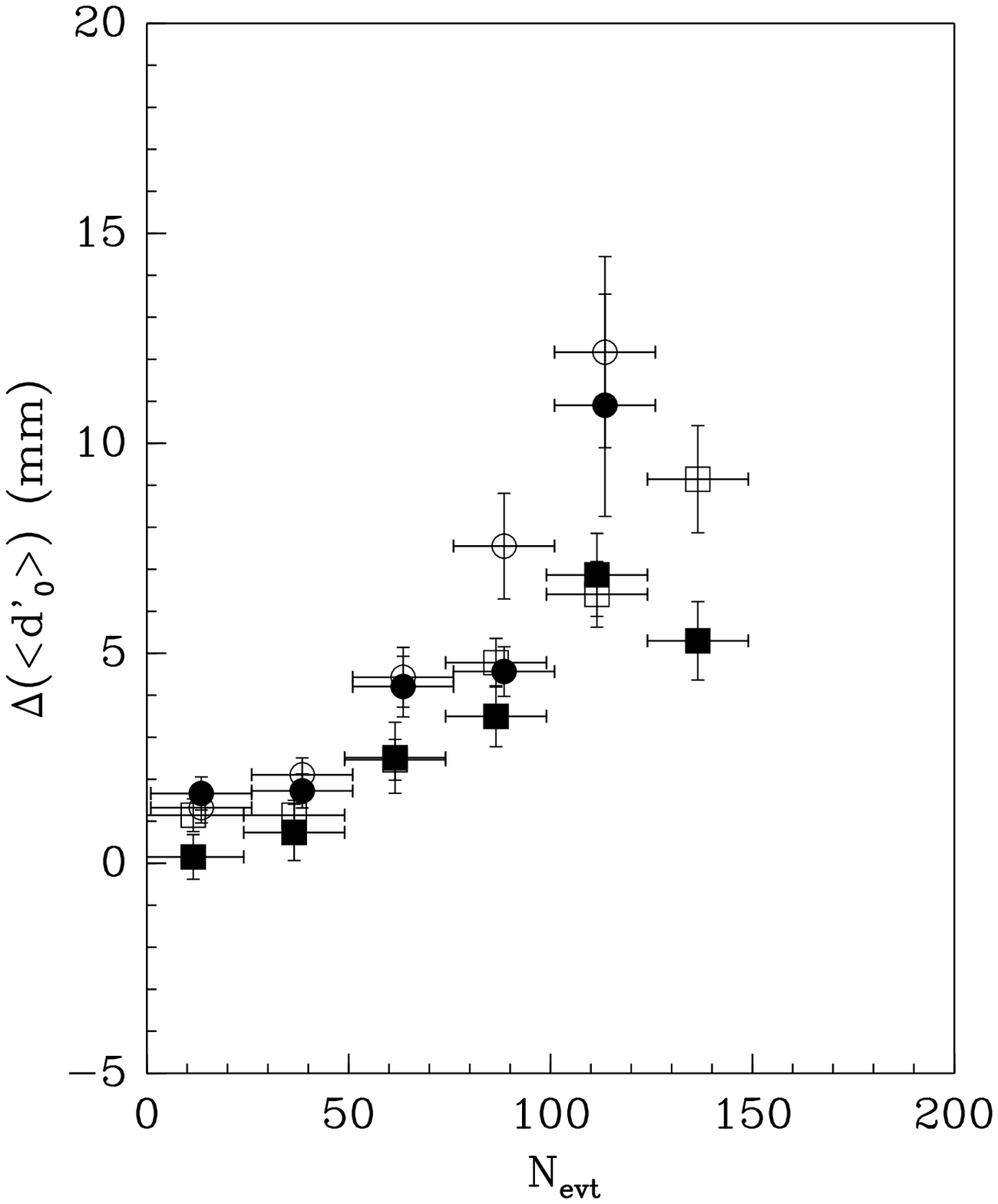}
  \end{center}
\caption{
Left panel: The shift in average momentum for  elastic scattering data (
3~\GeVc: open squares,  5~\GeVc: open circles) measured with elastic events as a
function of the momentum predicted by the forward scattered track. 
It is as function of the event number in spill for different predicted momentum.
The momentum
estimator from the fit not constrained by the impact point of the incoming beam
particle is used here. 
Right panel: The shift in average \dzeroprime as a function of the event number
in spill for  elastic scattering data ( 3~\GeVc: filled and open boxes, 5~\GeVc:
filled and open circles) measured with elastic events as a function of the
momentum predicted by the forward scattered track. The open symbols show the
data for momenta below 450~\MeVc and the filled symbols for momenta above
450~\MeVc.   
} 
\label{fig:el-p-bias-spill}
\end{figure}

By comparison with the momentum vector predicted with the elastic scattering
kinematics, it was verified with the data that the value of the polar
angle \tht is not modified by the dynamic distortions. 
However, the momentum measurement is affected. 
By disregarding the impact point of the incoming
beam particle during the fit, the curvature of the track in the TPC gas
volume can be measured directly.
One observes that the momentum and the  value of \dzeroprime are
biased as a function of \evtspill as shown in
Fig.~\ref{fig:el-p-bias-spill}.
The \evtspill dependence for \dzeroprime does not show a significant
difference between the lower and higher momentum part of the spectrum.  

The analysis of the elastic scattering
events sets very stringent constraints on the maximum effect of distortions of
all kinds on the measurements of kinematic quantities with the TPC. This method has provided
solid estimates of the systematic errors associated with distortions  
\emph{as a function of event-in-spill}. 
In particular, it has been used to estimate the overall systematic error on
the momentum determination~\cite{ref:tantalum},~\cite{ref:harp:MomCalib}.

\section{Ion Cloud Distortion Dynamics}
\label{sec:ion-cloud}

Given the beam intensity and the data acquisition rate with the 5\% interaction
length targets, it follows that HARP operated under conditions of
high dead time (higher than 90\%).
Hence, within one setting, the parameter \evtspill is a good measure of
the time the event was taken after the start of the spill.
(Depending on the precise data-taking conditions, the DAQ recorded about
one event per ms.) 

The beam instrumentation allows a precise evaluation of the direction, 
intensity and particle type of the impinging particles to be made. 
It is therefore
possible to show correlations of beam properties and dynamic distortions.
The beam is poorly focused during several settings\footnote{HARP uses the word
{\em setting} to define a group of runs with the same beam momentum and
polarity, target and trigger definition.}. 
It is observed that a badly tuned beam, with large number of halo particles 
which hit the target support material (representing several nuclear
interaction lengths $\lambda_{\mathrm{I}}$) 
induces a large number of particles during the data-acquisition dead time. 
The contrary also holds: dynamic distortions disappear or are strongly
reduced for settings where the beam is well focused, and can change from
run to run during a setting when the beam was re-tuned between these
runs.
Moreover, it is observed that the dependencies of the distortions  on the
azimuthal angle $\phi$ (observed for some settings) are correlated with
an off-axis tuning of the beam.

Beam particles hitting the beam entrance hole and the target support
material produce secondaries which enter the TPC and produce ionization charges.
Notice that a large number of low-energy (and therefore highly
ionizing) electrons are expected to enter the active gas volume. This kind
of phenomenon is likely to produce many more electron-ion pairs than the typical
triggered event in the target. 
The produced ionization electrons drift towards the amplification region
and then their number is multiplied near the 
pad plane with an amplification factor of the order of $10^5$, producing an
equivalent number of argon ions. 
Any inefficiency of the gating grid at the
level of $10^{-3}$ or even $10^{-4}$  allows an overwhelming number of
ions to reach the drift 
region and to start travelling in the TPC gas volume towards the cathode, forming
at the same time a positive charge cloud.
This charge cloud distorts the (otherwise uniform) drift field.

From measurements of positive ion mobility in argon based gas mixtures ~\cite{ref:alice},
the velocity of the ions is computed to be about 2~mm/ms (four orders of
magnitude lower than the velocity of drift electrons in the TPC). 

To help visualize the build-up and motion of the ion cloud, one should
keep in mind 
that the pad plane is at $z \approx -500$~\mm, the thin targets are at
$z \approx 0$~\mm, the
nose (Stesalite end disc) of the inner field cage is at about $z \approx
250$~\mm, and the total 
drift region is $\approx$~1500~\mm long (hence ending at $z \approx 1000$~\mm). 
Those numbers imply that, in a spill of 
about 400~\ms, 
the ion cloud just reaches the $z$ position of the target.

The constant increase of distortions during the spill is easily
explained: the drift electrons generated by tracks of triggered events
have to cross an increasing number of ions produced by the beam, as the
thickness of the ion cloud to be traversed increases from 0 to about 600~\mm
from the start to the end of a 400 ms long beam spill.
Before the period of linear increase in strength of the distortions, one
expects a short period of zero distortions: the ion cloud produced in the
amplification region first has to reach the drift volume.
This period is estimated to be about 25~\ms.
Between the period of zero effect and linear growth one expects a smooth
transition given by the ion diffusion and the difference 
in length of the drift path in the regions around the anode and gating 
grid wires.

\begin{figure}
\begin{center}
\includegraphics[width=0.49\textwidth]{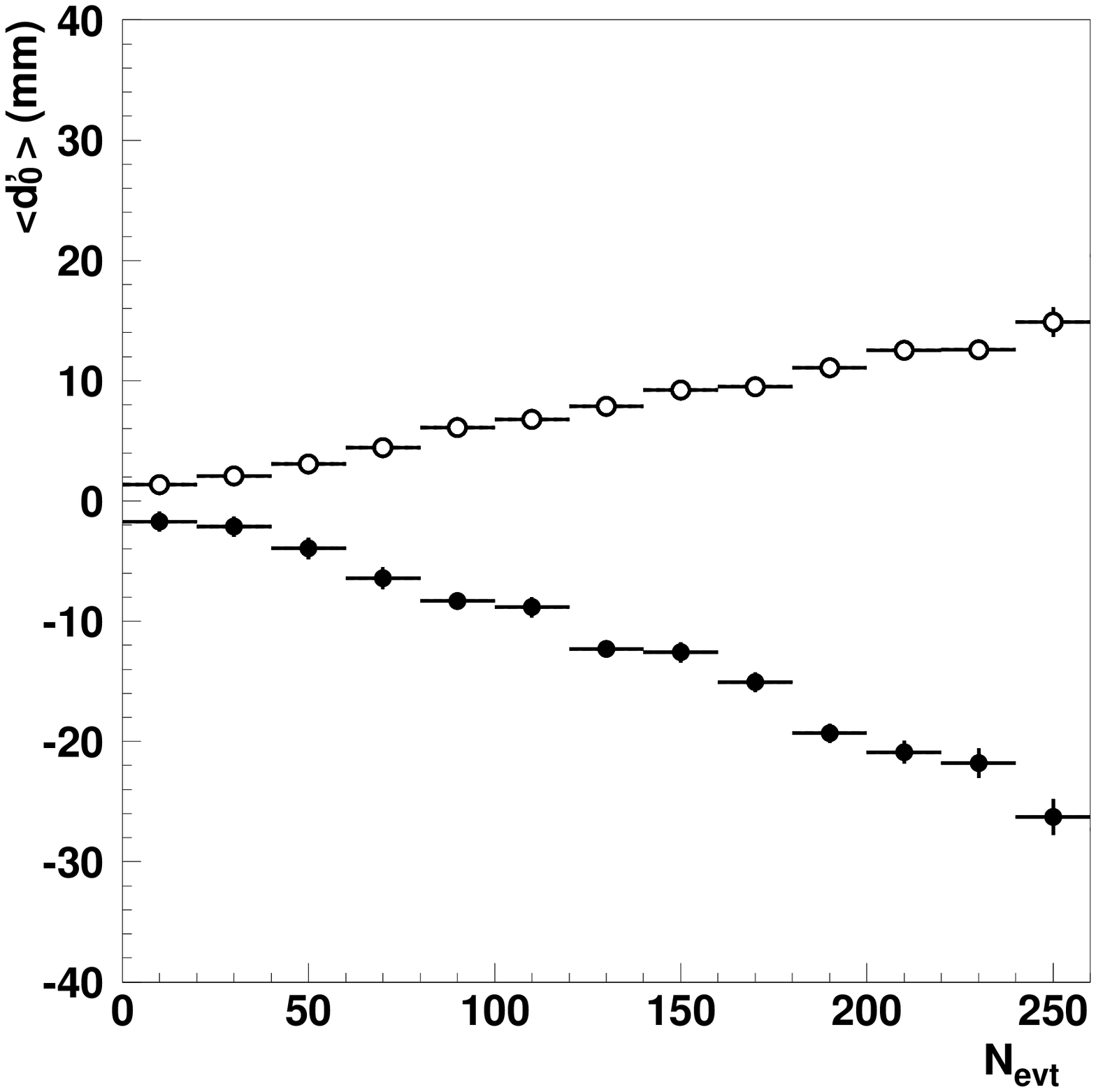}
\includegraphics[width=0.49\textwidth]{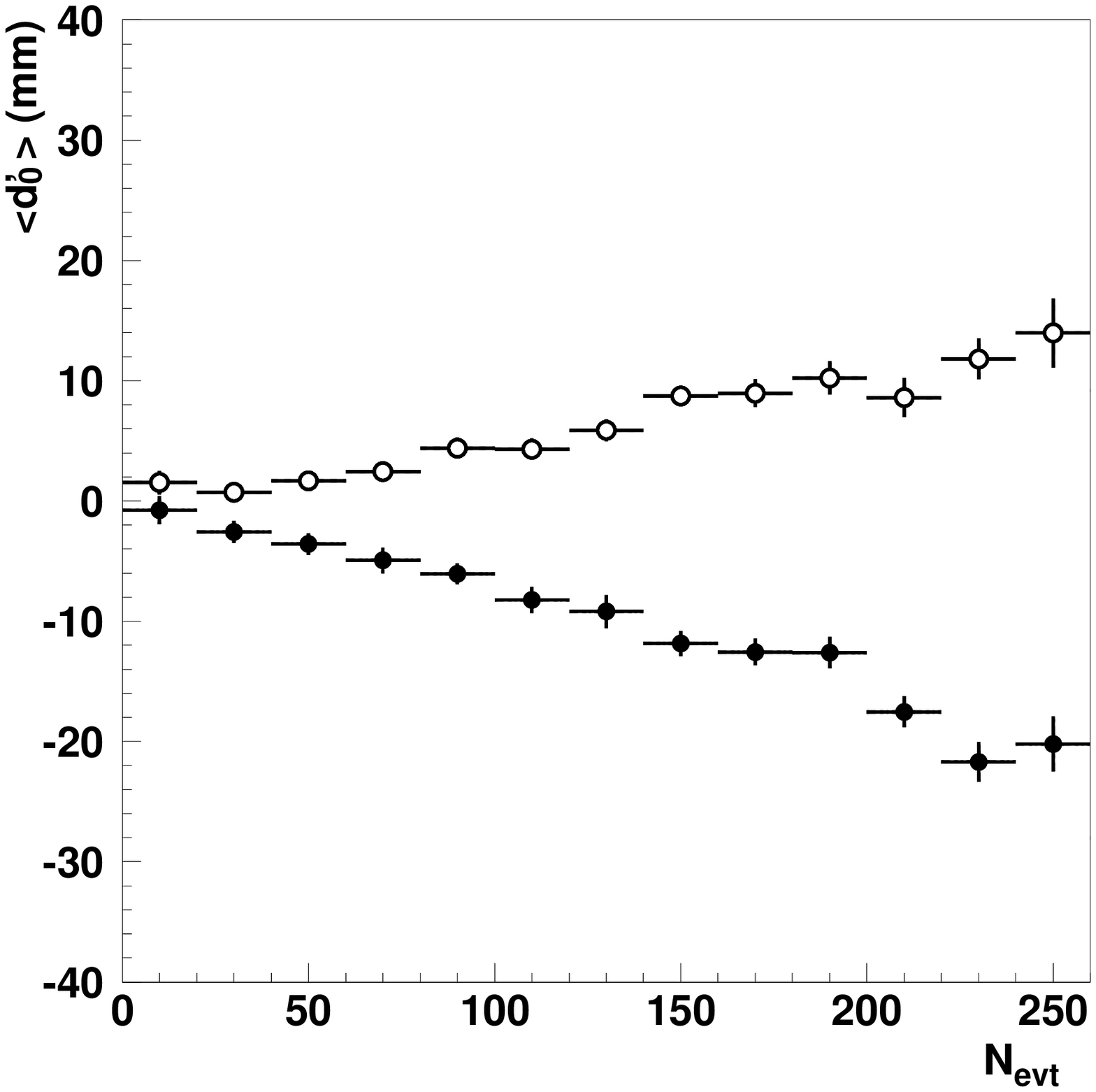}
\end{center}
\caption{The average \dzeroprime as a function of event number in spill 
for positively (open dots) and negatively (filled dots) charged particles
for the hydrogen target for 3~\GeVc beam. Left panel: distortion
of tracks originated in the target ($z \sim 0$ mm). Right panel: distortion of tracks
generated in the Stesalite ($z \sim 268.5$ mm).
Open circles: positively charged particles; closed circles: negatively
 charged particles.
}
\label{fig:distortion:all}
\end{figure}

These expectations can be tested by comparing the distortions affecting
tracks generated at different values of $z$ in the TPC.
Tracks within a limited angular range approximately perpendicular to the
beam direction are used for this analysis since their trajectory lies
within a small range in $z$. (Tracks between $\pm 30$ degrees with
respect to the normal are accepted.)
If the dynamics of the ion cloud is correctly described with the
considerations given above, tracks generated in the Stesalite end disc
of the IFC (large positive $z$) and in the
target should be affected by the same distortions at any time of the
spill, because their drift electrons have to cross the same ion cloud. 
Fig. \ref{fig:distortion:all} demonstrates that the distortions observed
for these two groups of tracks are indeed identical.

\begin{figure}
\begin{center}
\includegraphics[width=0.45\textwidth]{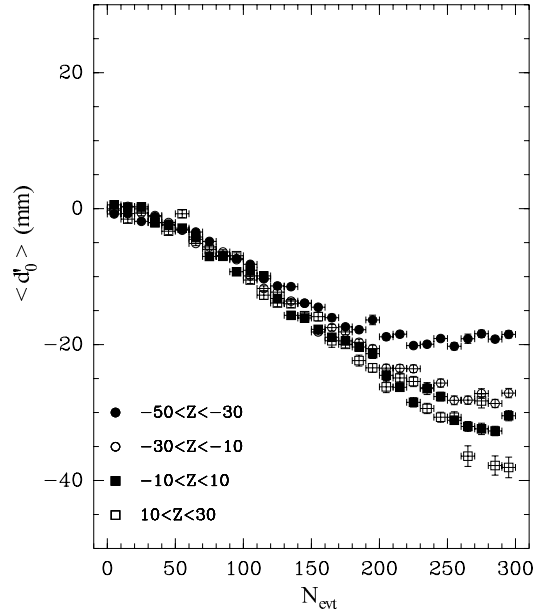}
\end{center}
\caption{
Average \dzeroprime measured as a function of the event number in spill
 using a long aluminium target.   
The data are divided into four regions of the $z$ of the interaction
 point. 
$z_0$: $-50 \ \cm \le z < -30 \ \cm$ (closed circles); 
$z_1$: $-30 \ \cm \le z < -10 \ \cm$ (open circles); 
$z_2$: $-10 \ \cm \le z < +10 \ \cm$ (closed squares); 
$z_3$: $+10 \ \cm \le z < +30 \ \cm$ (open squares). 
The data are shown for \pim tracks.
The event number in spill (in first approximation corresponding to time) where
the deviation
 of the average of the different  series of points saturate clearly show
 the ion mobility.
During the first $\approx$25 events (30~\ms) no deviation is visible, consistent
 with the fact that the ions -- created in the amplification region --
 have not yet reached the drift volume. 
During the following $\approx$50 events the derivative of the slope
 increases, showing the ion diffusion.
}
\label{fig:distortion:saturation}
\end{figure}

On the other hand, tracks generated in a long target (e.g. a two
interaction length long aluminum target of $\approx$80~\cm) at
negative $z$ values in the TPC (half-way between the pad plane and $z=0$), should
show a saturation of the distortions before the end of spill. 
Figure~\ref{fig:distortion:saturation} indeed shows that the distortion of tracks
produced at $z \approx -250$~\mm is no longer increasing after about 130~\ms,
consistent with the predictions given in Section~\ref{sec:ion-cloud}.
The ion wavefront is expected to reach $z \approx -250$ mm after about 125~ms, and
the thickness of ion cloud to be traversed by the drift electrons remains
constant.
(Given the rather uniform beam intensity during the spill, per unit time
the same number of  ions are produced at the pad plane as the  
number which cross the ideal $z = -250$ mm plane.) 
Fig.~\ref{fig:distortion:saturation} also shows that tracks produced at increasingly
larger $z$ exhibit the distortion saturation at increasingly later
times.

\begin{figure}
\begin{center}
\includegraphics[width=0.45\linewidth]{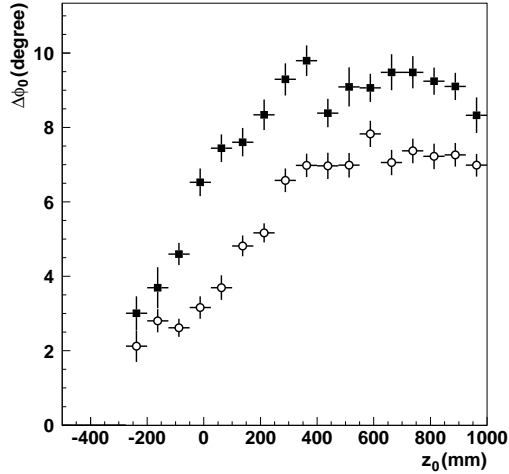}
\end{center}
\caption{
The average $\Delta \phi_0$ for two time
periods with different delays from the end of the preceding spill.
Cosmic rays taken within the first 175~\ms after the end of the spill
 are shown with closed circles, while the tracks taken between 250~\ms
 and 300~\ms  after the end of the spill are shown with open circles.
The distortions tend to zero at $z$ values which are already passed by
the ion packet.}
\label{fig:distortion:interspill}
\end{figure}

Ions are no longer generated in the amplification region after the end
of the beam spill.
Thus it is expected that the ion cloud remains of constant thickness
(about 600~\mm) between spills, and that it drifts into the direction of
positive $z$, gradually freeing the active volume from 
distortions starting first with the negative $z$ region of the TPC. 
Cosmic-ray tracks recorded for calibration taken during the time between
spills allow this behaviour to be studied.
To be able to study the distortion effects, tracks are selected which
approximately cross the IFC region.
The distortions in the measurements of the trajectory on either side of
the region of the IFC are of opposite sign (if expressed in Cartesian coordinates).
As a measure of the distortions the variable $\Delta \phi^0$ is defined
as the difference of the measured $\phi$ of the top half of the track
compared to the $\phi$ measured for the complete cosmic-ray track.
Figure~\ref{fig:distortion:interspill} shows the $\Delta \phi^0$ for two time
periods with different delays from the end of the preceding spill.
Indeed the two back-to-back segments of cosmic-ray tracks taken at
negative $z$ become progressively less affected by distortions, and
the distortion-free region expands with time (while 
the ions drift towards more positive $z$).

\begin{figure}
\begin{center}
\includegraphics[width=0.45\linewidth]{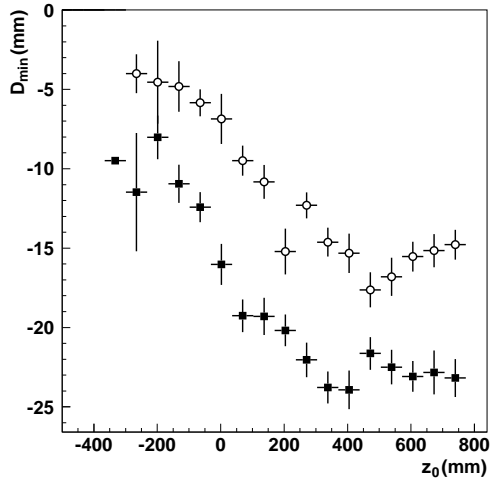}
\end{center}
\caption{The dependence of the distortion for cosmic-ray tracks at large $z$
on the position of ion cloud during the inter-spill period.
{\em Dmin} is the minimum distance of the trajectory of one arm of the
 cosmic-ray track with respect to the point closest to the origin
 evaluated for the complete  cosmic-ray track. 
Cosmic rays taken within the first 175~\ms after the end of the spill
 are shown with closed circles, while the tracks taken between 250~\ms
 and 300~\ms  after the end of the spill are shown with open circles.
}
\label{fig:distortion:interspill_largeZ}
\end{figure}

Subtle effects can be shown and explained by analyzing the cosmic-rays
taken between spills.
From Fig.~\ref{fig:distortion:interspill_largeZ} it can be observed that
tracks at very large positive 
$z\ge 550$~\mm see a different distortion strength at the beginning of the
inter-spill (first cosmic) with respect to the distortion seen after about
additional 80~\ms--100~\ms after the end of the preceding spill.
This occurs despite the fact that in
both cases their drift electrons cross the same ion cloud
thickness.
The front of the cloud has meanwhile just moved by less than 200~\mm,
and did not reach yet $z=500$~\mm. 
This holds true even if taking into 
account that the trigger system was programmed to provide a wait-state of about
140~\ms between the end-of-spill and the first inter-spill cosmic. 
The reasons for this behaviour can be fully understood: during the 80~\ms--100~\ms
difference between the first and last inter-spill cosmic-rays, a fraction of the ion
cloud has passed the position of the inner field-cage and disc, entering a region
where the electrostatic configuration of the field-cage is completely different.
In the region where the electric field is only formed by the cylinder of the
outer field-cage, the distortion field produced by the ions in the disc
with $R$ smaller than the drifting electron clusters
is only attractive towards the origin: the repulsion term of the inner field cage is
missing, and the outward component of the ions at radii larger than the
position of each drift electron is null. 
Therefore, there is a range in $z$ in which
the the drift electrons feel an inward force, thus partially
compensating the usual distortion at small R in the $z$ range, where both the ions and the
inner field cage are present. This is why the cosmic-ray tracks crossing the TPC
at large $z$ values are more distorted at times directly after the spill
compared to later times.
There is a $z$ value (taking in account the trigger shift), corresponding to the 
end/cap of inner field cage, where the distortions have a maximum, see 
Fig.~\ref{fig:distortion:interspill_largeZ}.

\section{Experimental determination of the $R\phi$ distortion using elastic
scattering}
\label{sec:elastics-shift}

When trying to measure the effect of distortion in an unbiased way it is
important to avoid the use of reference quantities which can themselves
be affected by the distortions.
For example, a biased result is obtained if one measures first the track
curvature using the distorted trajectory and if one then at a later
stage uses this curvature in combination with fixed references such as
vertex position and hits of RPCs to predict the undistorted position of
charge clusters inside the gas volume.
This approach was suggested in  \cite{ref:tpc:dydak}  and we show in the
following that such a procedure can be avoided.

On the contrary,
as already discussed in Sec.~\ref{sec:elastics-effect}, 
elastic scattering off H$_2$ can be used to predict the complete
undistorted trajectory without making use of quantities which are
affected in any way by the distortions.
By measuring the scattering angle ($\theta$) of the forward going particle
with respect the direction of the beam particle (whose momentum is precisely
selected by the beam setting) the four-momentum of the proton recoiling at 
large angle is derived from the elastic scattering kinematics.
This provides a reference quantity suitable to actually {\em measure} 
the distortion.  
The knowledge of the four-momentum of the large-angle proton is the key
to extend the method to directly determine the $R\phi$ displacement of
the clusters.  
This approach avoids completely the introduction of dependencies on
parameters affected by the distortions.

The full trajectory of the large-angle proton in the active region of
the TPC is calculated by using the geometry of the detector as
described in detail in the simulation program. 
The simulation program takes into account all the details of the
materials traversed by the scattered proton. 
This creates for every individual pad row an \emph{unbiased} reference
sample as function of \evtspill free from {\em a-priori} assumptions. 

\begin{figure}
\begin{center}
\includegraphics[width=0.45\linewidth]{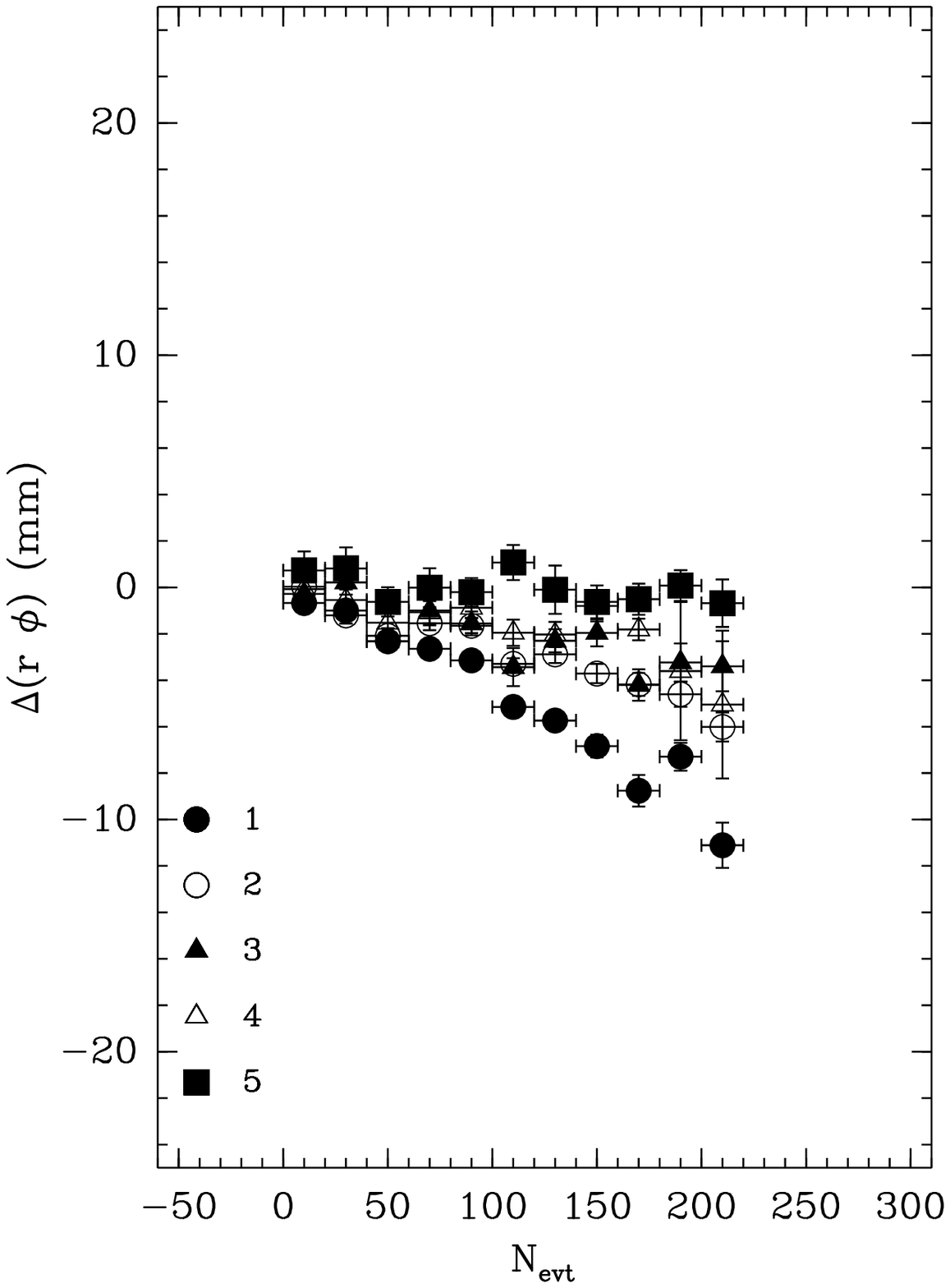}
\includegraphics[width=0.45\linewidth]{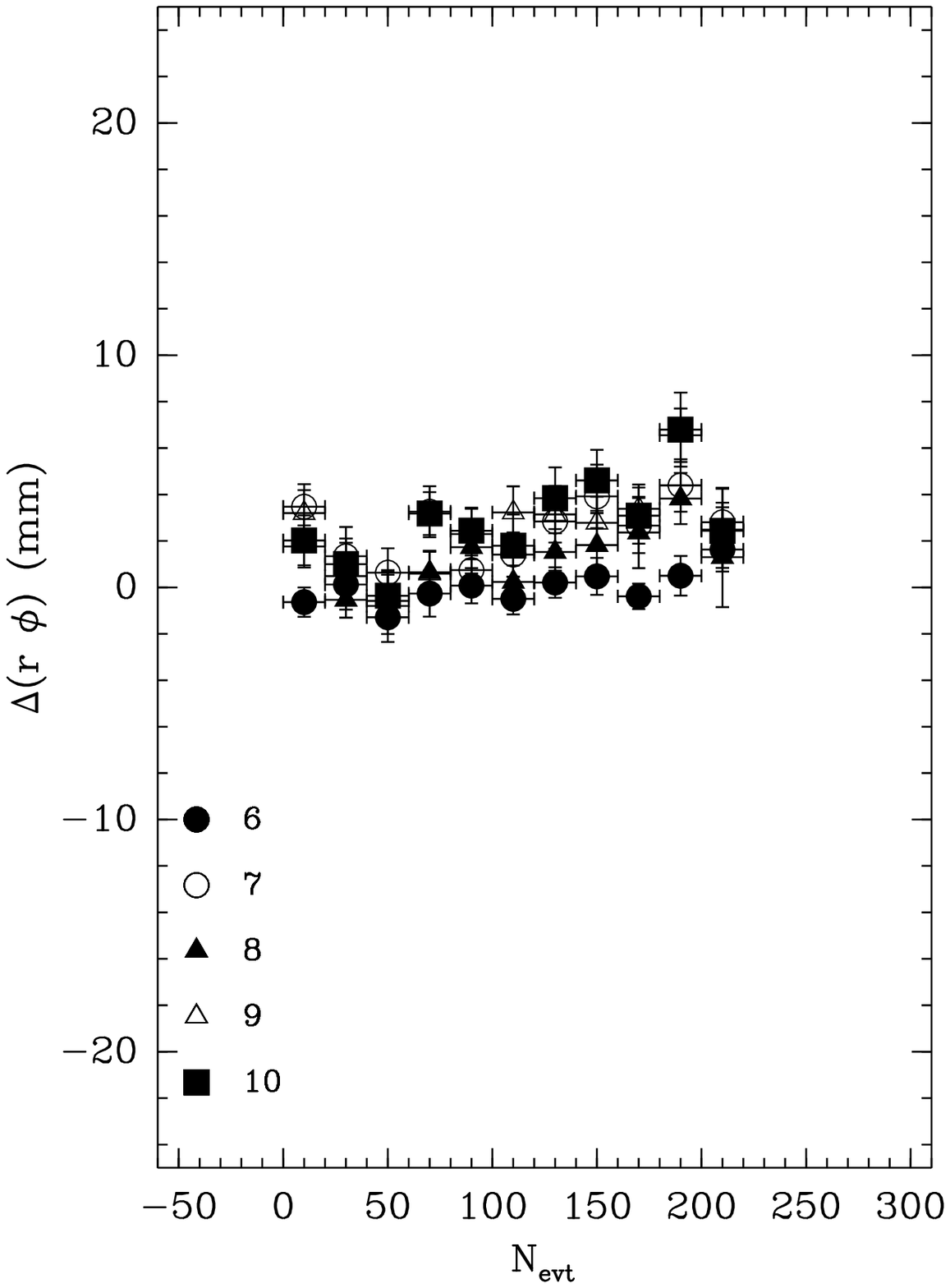}
\includegraphics[width=0.45\linewidth]{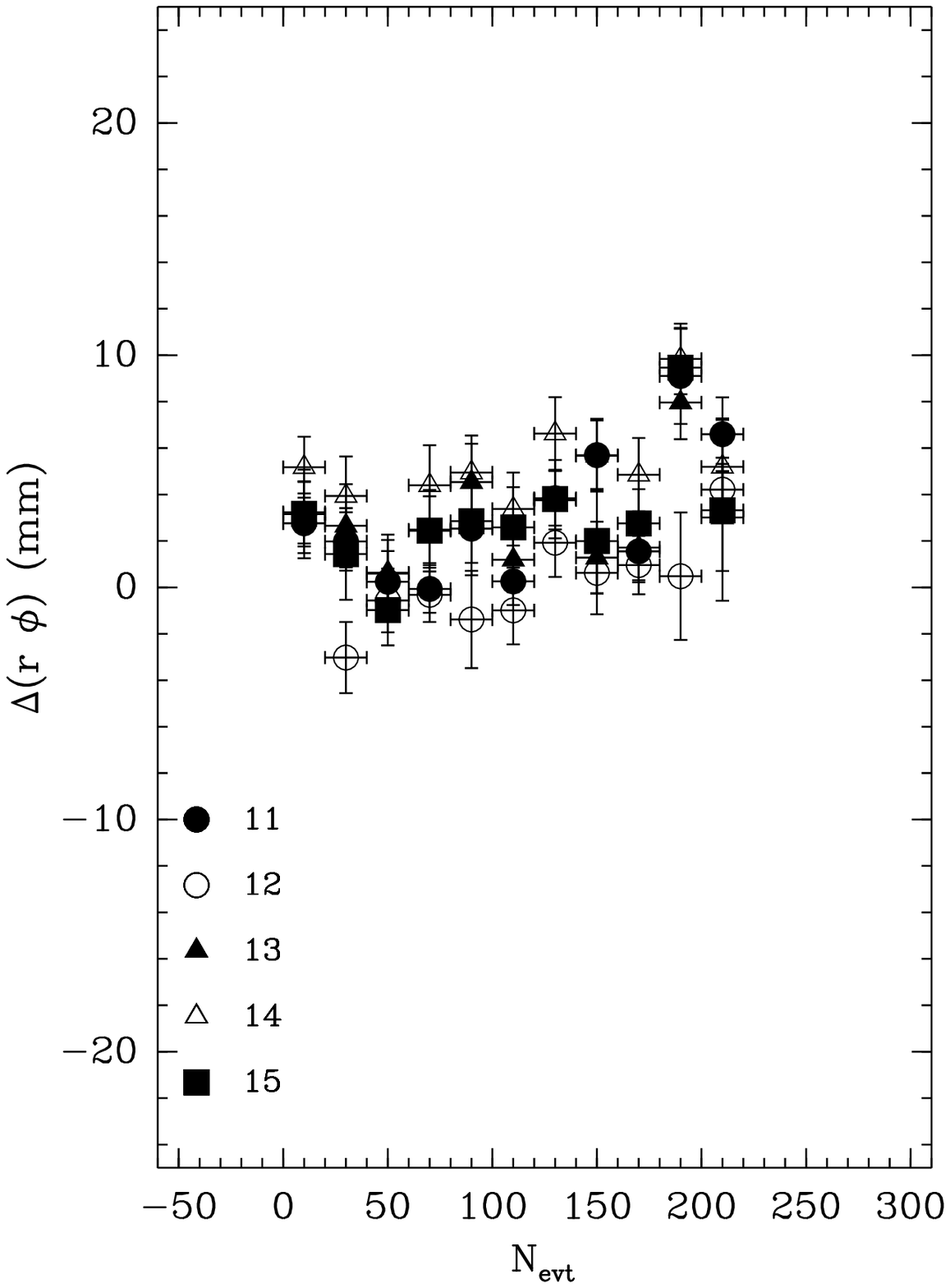}
\includegraphics[width=0.45\linewidth]{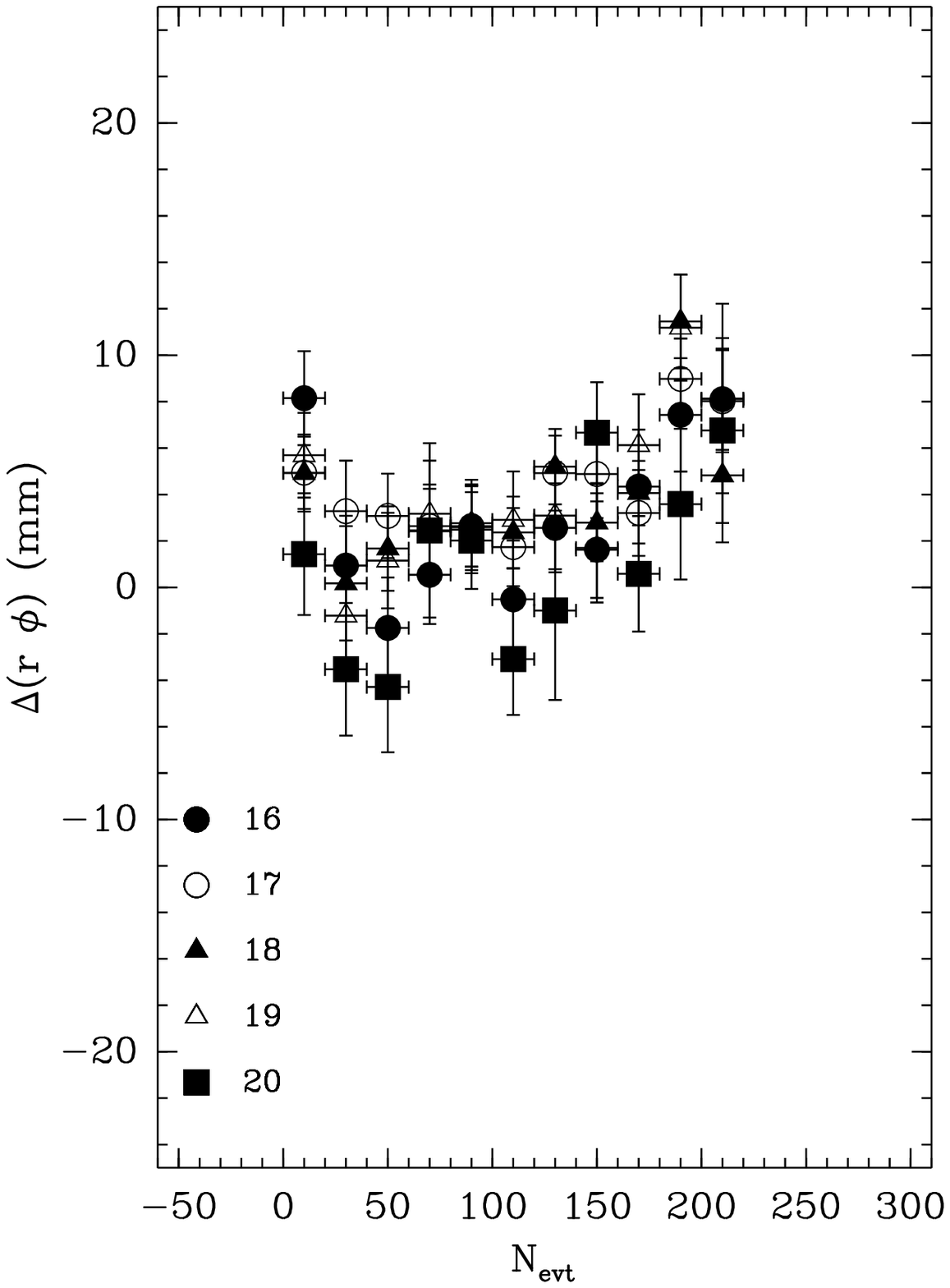}
\end{center}
\caption{The $R\phi$ distortion $\Delta (r \phi) = \langle(r \phi)_\mathrm{measured}
 - (r \phi)_\mathrm{predicted}\rangle$
 measured row--by--row as a function of event number in spill for the 3 \GeVc beam
and 180 cm long H$_2$ target data.
The four panels show data for four groups of five pad-rows each.
The different symbols represent the individual pad-rows.
}
\label{fig:rphi-measured}
\end{figure}

\begin{figure}
\begin{center}
\includegraphics[width=0.45\linewidth]{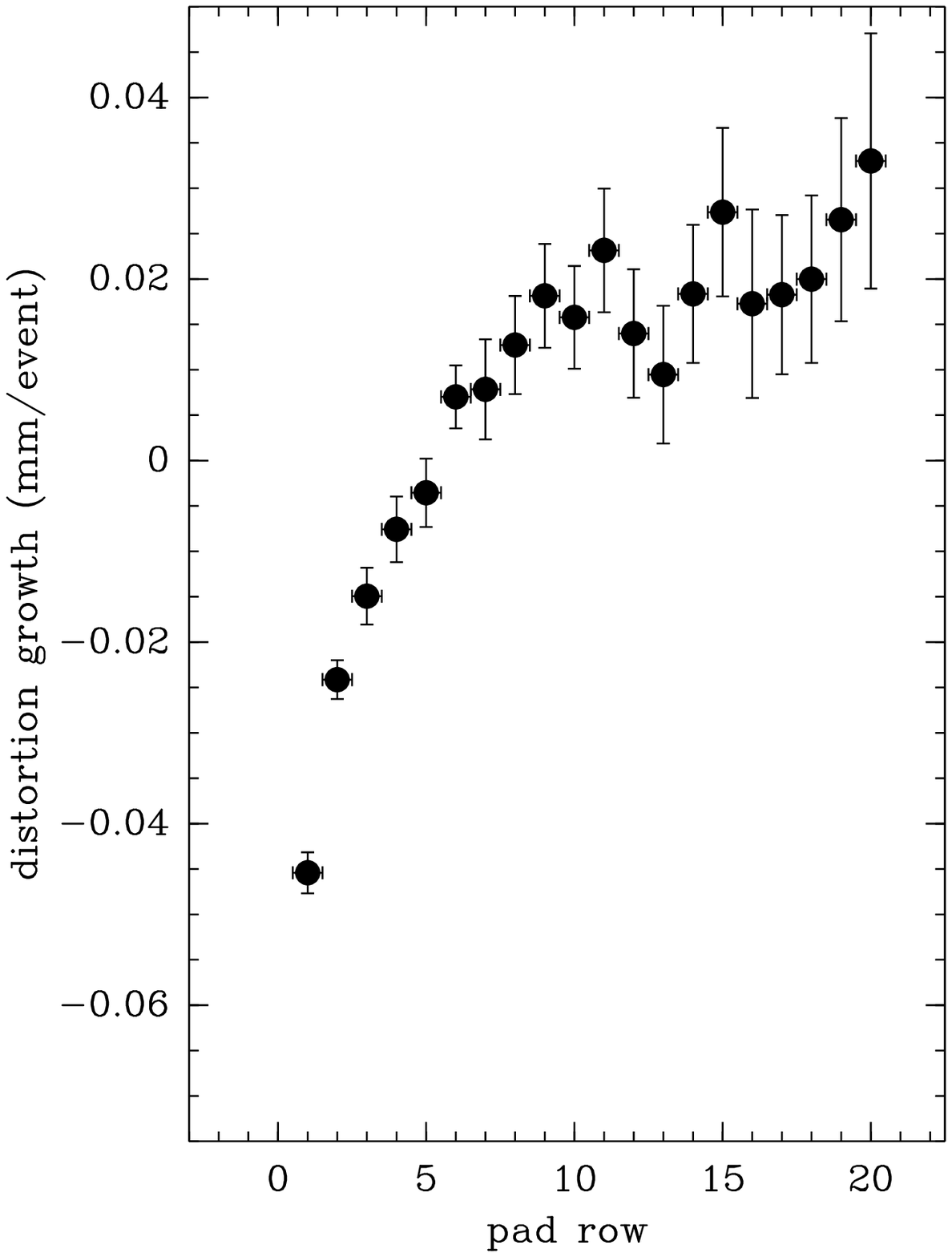}
\includegraphics[width=0.45\linewidth]{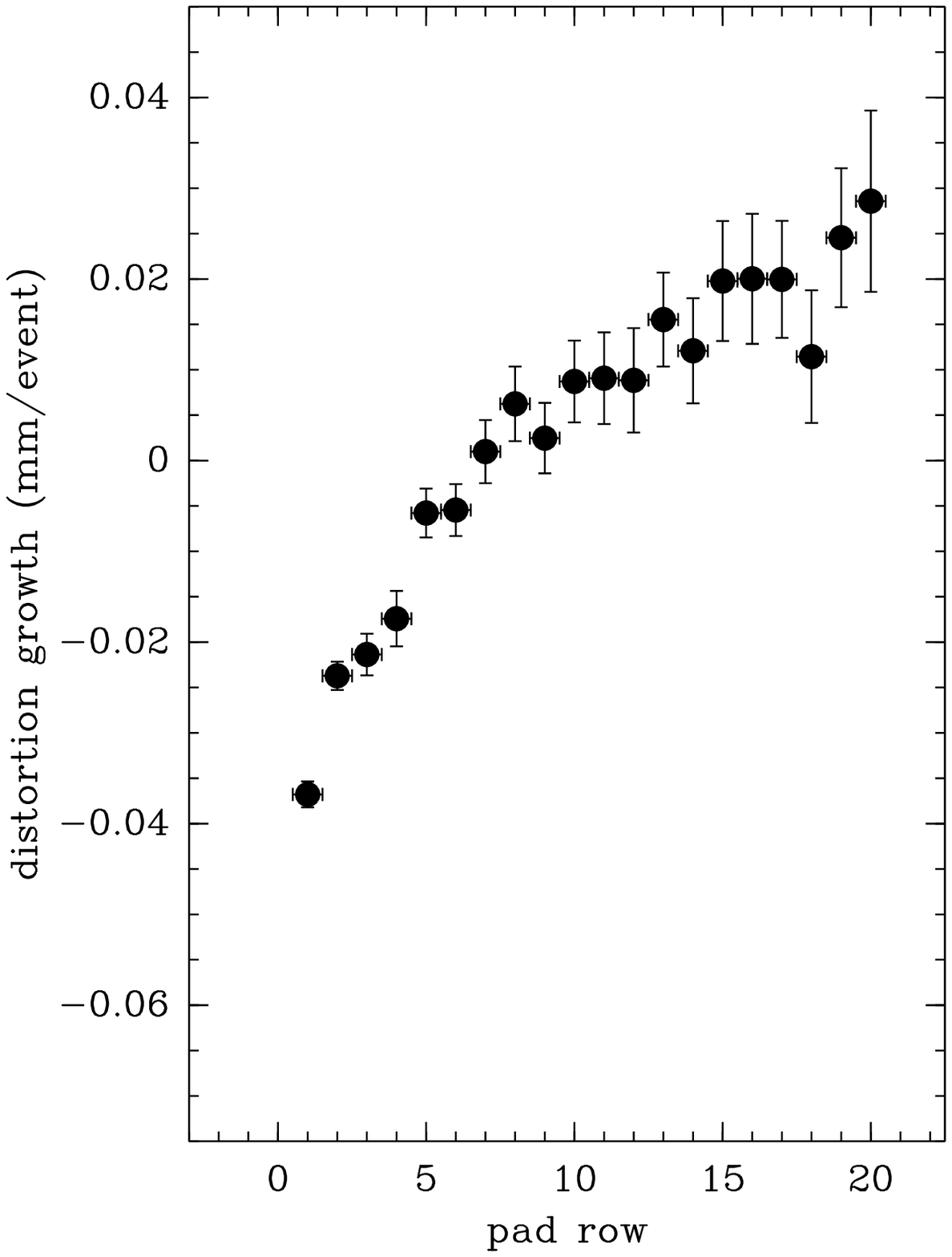}
\end{center}
\caption{The variation of the distortion strength as a function of the
 pad row number, fitted during the whole spill, left
 panel: 180~\mm H$_2$ target, 3~\GeVc beam; right
 panel: 60~\mm H$_2$ target, 5~\GeVc beam. }
\label{fig:distortion:slopes}
\end{figure}

The procedure was applied to the five reference hydrogen data sets
available: 3~\GeVc,  5~\GeVc and  8~\GeVc with a 60~\mm long target and 
3~\GeVc and  8~\GeVc with a 180~\mm long target. 
The average difference (along $r\phi$) of the position of the predicted
trajectory and the measured $r\phi$ coordinate 
are shown in Fig.~\ref{fig:rphi-measured} as a function of \evtspill 
for data taken with the 180~\mm hydrogen target in the 3~\GeVc beam.
For each pad plane row a straight line fit of the distortion measurements 
during the whole spill is made. The slope of the best straight line fit 
is used as monitor of the growth of the distortion versus time, is called 
distortion strength, and is given in units of growth of the distortion 
per recorded event.
Figure~\ref{fig:distortion:slopes} shows the results obtained for the
3~\GeVc beam impinging on the 180~\mm target and the  5~\GeVc data taken
with the 60~\mm H$_2$ target.
The distortion strength increases during the whole
spill, consistent with the behaviour of \dzeroprime  shown in fig. 
\ref{fig:distortion:all}.
Most interestingly, Fig.~\ref{fig:distortion:slopes} also shows that the
direction of the
distortion changes sign from the inner TPC rows to the outer ones, and
that there is a cross-over point of vanishing distortion.
The change of sign can be explained qualitatively with electrostatic
arguments  taking into account the fact that the HV power supply keeps
the inner and outer field cages at a constant voltage.
These arguments will be worked out in detail below.
One can further observe that the absolute value of the outward field
component at row number one is larger than the absolute value of the
inward field component at row number twenty.

\section{Phenomenological Model}
\subsection{Simple discrete model}
A phenomenological model can be constructed  based on the fact that the
field which is responsible for the force acting on each drift electron
is equivalent to a vector sum of two field systems: 
\begin{itemize}
\item{} a field system where ions, in a given angular section at $R$ values
        internal to the drift electron position contribute to attract the drift 
        electrons inward;
\item{} a field system where ions, in a given angular section at $R$ values
        external to the drift electron position, contribute to attract the drift
        electrons outward.
\end{itemize}

\begin{figure}
\begin{center}
\includegraphics[width=0.9\textwidth,angle=0]{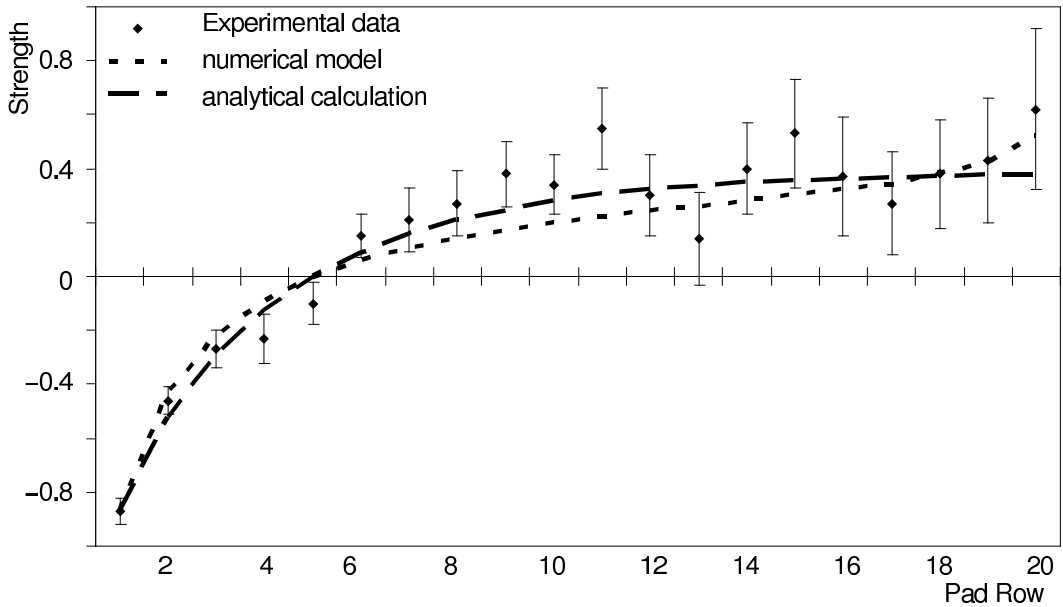}
\end{center}
\caption{
The comparison of the measured distortion and the analytical models.
The strength is shown with an arbitrary scale.
Closed diamonds are the experimental data; long dashes show the analytical calculation 
for uniform space charge Eq.~(\ref{eq7}) and  short dashes the numerical 
model.
}
\label{fig10}
\end{figure}

In this description, the fixed voltage of the inner field cage plays a
crucial role: it breaks the circular symmetry by shielding the charges
in its shadow.
This model makes it possible to understand all the peculiar features of the TPC dynamic 
distortions:

A simple, approximate algorithm can be written, using a discrete
representation at the row-level.
Despite of its simplicity such a model can predict the basic features
with surprising accuracy.
Various models of the distribution of the density of ions as a function
of $R$ and of the R dependence  of the electric field can be readily
modelled.
These models predict the change of sign of the  distortion strength
between the first and last pad rows, as well as the position of the pad row with
vanishing distortion.
The system can be modelled with only a very small number of fixed geometrical
parameters of the TPC and a free parameter describing the overall
distortion strength.  
A comparison
with the distortion measured with the data taken with the  180~\mm
hydrogen target exposed to the 3~\GeVc beam is shown in Fig.~\ref{fig10}. 

\subsection{General Analytical Solution}

The radial electric field distorting the trajectory of the drift electron is due
to two components. Given any radius $r_e$ where a drift electron is supposed to
travel parallel to $z$ in the TPC, a distortion term is directly due to the
contribution of the positive ion cloud integrated from the first row up to
$r_e$. The cylindrical shells of the ion cloud, external to $r_e$, do not
contribute because of the Gauss theorem. A second distortion term is due to the
induced excess of negative charge onto the conducting surface of the inner field
cage. Such a field is equivalent to the one of a charged wire along the $z$ axis.

The two distortion terms have opposite sign. The repulsive (outward force)
term prevails at small $R$. The attractive (inward force, same sign as the
static distortion force) term prevails at large $R$. There must be a radial
position with vanishing distortion effect. 
A simple discretization of the field due to the ions alone, at the
radius $R=r_e$, can be written as:

\begin{equation}
E_r^1(R)=\frac{2\sum_{i}Q(R_i)}{ZR},\;\;R_i<R \ ,
\label{eq1}
\end{equation}   

because the field due to each discrete cylindrical shell is equivalent to the
one generated by a wire length $Z$ charged with the charge $Q(R_i)$ contained in
the volume of each shell. 
R is the radius at which the field has to be computed, which is at the same distance
from all effective wires along $z$ {\it generated} by each cylindrical shell, according
to the Gauss theorem with the assumption of uniform charge density along $z$ and over 
the cylindrical angle.

The ions are generated by the amplification of all the ionization electrons 
produced by all the charged particles that traverse the TPC during the spill.
The ion cloud in the drift volume is populated by ions which cross the grid in front 
of the anode wires either because the grid was open or not perfectly closed.
Interactions of beam and halo particles with the target and its surrounding materials
produce secondary particles most of each reach the outer field cage of the TPC.
Therefore, a $1/r$ dependence of the ion charge distribution in space is
a very good initial estimate. 
This consideration gives:

\begin{equation}
E_r^1(R)=\frac{K_1}{R}\int_{R_1}^{R}{\frac{dr}{r + c}} \ ,
\label{eq2}
\end{equation}   

where $K_1$ is a charge normalization factor depending on the beam and
target setting, $c$ is
the parameter fixing the charge at any given radius (also depending on
beam and target setting), $R_1$ is the internal radius of the ion cloud
(close to the  first pad row).

On the other hand, the field due to the induced negative charge $-Q_2$ on the
inner field cage is given by:

\begin{equation}
E_r^2(R)=-\frac{K_2Q_2}{R} \ .
\label{eq3}
\end{equation}   

This relation is correct as long the inner field cage is surrounded by the
ion cloud, as it is indeed the case, because the ion velocity is such that the
ion wavefront reaches the target position $z = 0$ by the end of the
spill. 
Thus the total radial distortion field can be expressed as:

\begin{equation}
E_r(R)=E_r^1+E_r^2=\frac{1}{R}\left(K_1\ln \left(\frac{R +
c}{R_1+c}\right)-K_2Q_2\right) \ .
\label{eq4}
\end{equation}   

The general expression can finally be written as:

\begin{equation}
E_r(R)=\frac{K}{R}\left(\ln \left(R + c \right)-y \right) \ ,
\label{eq5}
\end{equation}   

where $K$ and $y$ absorb the various constant factors. 
This equation contains the setting dependencies of the charge
distribution and beam intensity in the form of the free parameters $K$,
$y$ and $c$ and they can be used to parametrize and correct the 
dynamic distortions of any setting.
The number of parameters can be reduced by the condition:

\begin{equation}
\int_{R_1}^{R_2}E_r(R)dr=0 \ ,
\label{eq6}
\end{equation}   

where  $R_1$ is  the conducting surface of the inner field cage and $R_2$ is 
the conducting surface of the outer field cage. 
This condition uses a natural assumption that the power supply manages to keep 
the voltage at the nominal value at each resistor partition
(we assume here that the static distortions have been corrected).

The absolute scale to be used in Eq.~\ref{eq5} can be fixed with the
help of experimental data.
In the hydrogen settings, the direct measurement of the position of the
clusters can be used, while in the data taken with the other targets the
time-dependence of the distribution of $\langle \dzeroprime \rangle$ is
a good estimator.

Assuming a uniform distribution of the ion charge in the volume,
Eq.~\ref{eq6} can be resolved analytically if one requires $c \ll R$ 
(justified by the charge generation mechanism around the beam), and if one neglects 
the difference between inner field cage radius and the inner radius of the 
ion cloud (which is in any case internal to the first pad ring radius) 
Then  this gives a solution for the value of $y$:

\begin{equation}
y=\ln \left(R_0\right) \  \mbox{ where } R_0 = \sqrt{R_1R_2}\,
\label{eq7}
\end{equation}   

{\em i.e.} $R_0$ is the geometric average of the radial positions of the
conducting surfaces and is the radius for which the resulting field is zero. 
Thus this could be predicted from first principles. 

A detailed comparison with the measurements is shown in
Fig. \ref{fig10}, using Eq.~\ref{eq5} and \ref{eq7} and 
the geometric dimensions of the TPC surfaces and pad size. This shows a
remarkable agreement between the distortion data for the 3~\GeVc hydrogen data
and the prediction based on the uniform charge distribution.
If one uses a non-uniform space charge distribution the
numerical solution of Eq.~\ref{eq6} gives similar results.

Both with the numerical integration code and with the analytical calculation, the
electrostatic problem has been solved also for different density distributions
(and different $R$-dependencies of the  electric field taking into
account more or less strong edge effects). 
This may be useful in case of particular beam settings.
For example, below is given the formula expressing the radial electric field for
the case of a $1/R$ superficial density distribution on each cylindrical shell
(i.e. constant charge in the volume of each cylindrical shell):

\begin{equation}
E_r(R)=E_0 \left(1 -\frac{R_2 - R_1}{R\ln(R/R_1)} \right) \ .
\label{eq8}
\end{equation}   

We have shown that the method works for other density distributions; however the
density distribution used in the paper is the one resulting from the raw
measurements of the pad occupancies.

\section{Correction method}
\label{sec:correction}

With the models described above and the direct measurements a
\emph{distortion strength} as a function of row number is determined. 
The \emph{strength} is measured as a residual, therefore it can be used
as the basis for an $R\phi$ correction applied to clusters measured on
tracks.  
However, before the correction can be used for each target and beam
setting where the elastic scattering cannot be measured, one has to correlate
the characterization with the behaviour of one or more global
track parameters. 
The analysis of the elastic scattering data shows that the largest
effect of the distortions is seen in the pad rows nearest to the
centre. 
Therefore, one of the best candidates in this respect is \dzeroprime
which is easy to measure for each track and therefore gives
statistically significant reasults for each data set. 

\begin{figure}
\begin{center}
\includegraphics[width=0.43\textwidth,angle=0]{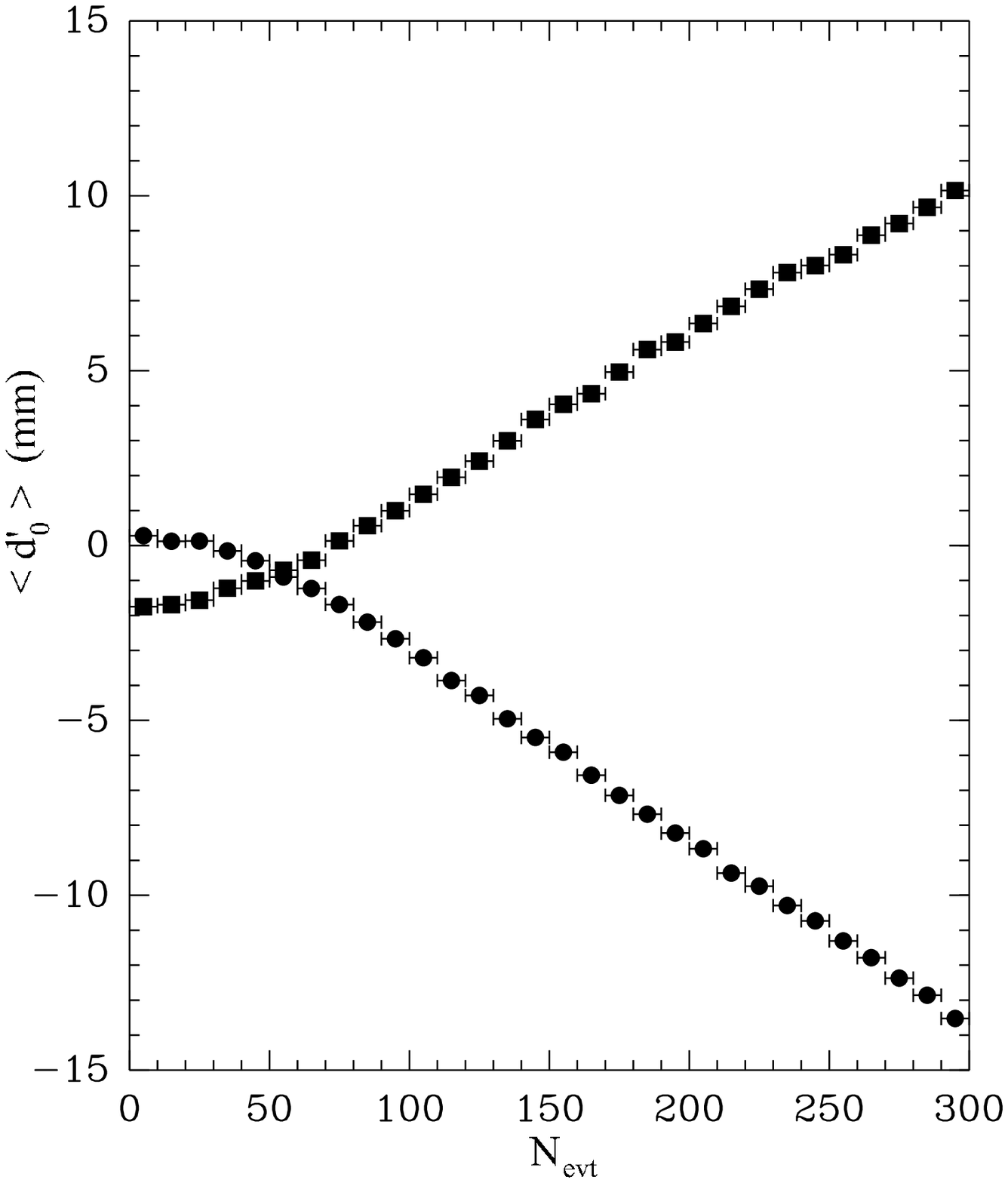}
~
\includegraphics[width=0.43\textwidth,angle=0]{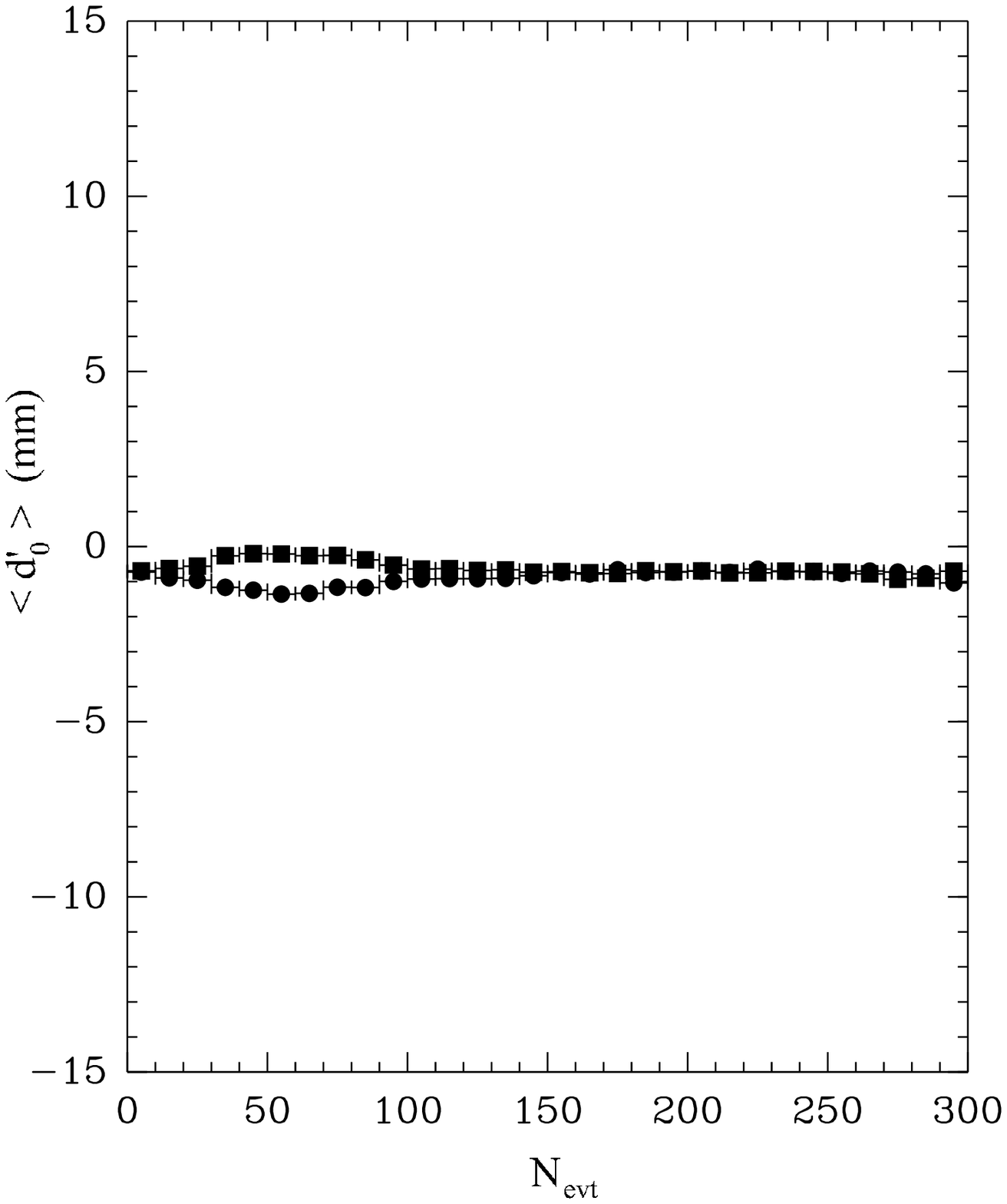}
\end{center}
\caption{
Average \dzeroprime (dots for reconstructed positive tracks, squares for
reconstructed negative tracks) as a function of event number in spill for 8.9~\GeVc
 Be data. (left panel uncorrected; right panel: dynamic distortion
 corrections applied.)
After the ``default'' correction for the static distortions (equal for
 each setting) a small residual effect at the beginning of the spill is
 visible at $\evtspill=0$ (left panel).
This is due to the fact that the inner and outer field cages are powered
 with individual HV supplies.
A setting-by-setting correction compatible with the reproducibility of
 the power supplies is applied for the data of the right panel together
 with the dynamic distortion correction.
The value of $\langle \dzeroprime \rangle$ at $N_{evt}=0$ in the right panel has a small negative
value as expected from the fact that the energy-loss is not described in the 
track-model used in the fit. The difference observed in the results for the
two charges around $N_{evt}=50$ shows that even at the onset of the effect the 
model can correct the distortion within 1 mm.
}
\label{fig:be:dzeroprime}
\end{figure}

The behaviour of \dzeroprime as a function of the event-in-spill shows
a first part with, essentially, no distortion, then a quadratic rise, followed
by a linear behaviour until an upper limit is reached. Empirically, this
behaviour can be understood from the previously described ion cloud dynamics.
At the beginning of the spill the TPC is distortion-free; soon after the 
onset of the distortion the effect stabilizes into a linear increase
until levelling-off at the point in time where all the distance
travelled by the ionization charges is filled by the ion cloud.
The saturation is in practice not reached for tracks emanating from the
target during the spill.
The intermediate region approxamated by the quadratic rise is understood
as the onset of the effect when the front of the ion cloud enters in the
drift region.
Due to the different paths the ions travel in the amplification region,
the front of the cloud is not sharp.
The behaviour is seen in Fig.~\ref{fig:be:dzeroprime} (left panel).
It is observed that the dependence of \dzeroprime as a function of 
\evtspill shows the three regimes of the distortion as described above. 
Three calibration parameters are extracted with an iterative
procedure: the value of \evtspill up to which there is no distortion,
the value of \evtspill where the rise changes from quadratic to
linear, and an overall scale factor. 
A time-dependent upper limit to the growth is also defined 
to take into account the fact that the distortion saturates at a different
value of \evtspill depending on the $z$ position of the original
ionization charge.
This is not a free parameter. 

To take into account the different characteristics of the initial charge
distribution the data taken with the 3~\GeVc beam use the corrections
determined using the  3~\GeVc hydrogen data, the 5~\GeVc beam
corrections use the 5~\GeVc hydrogen data, while the 8~\GeVc hydrogen
data are used for the higher momenta.

For a given setting all data are first reconstructed without any
correction for dynamic distortions and with a default
(setting-independent) correction for static distortions.
The characteristics of the dependence of \dzeroprime on \evtspill is
then used to determine the initial values for the four parameters (three
for the dynamic distortions and one for the static correction.)
The row-by-row dependence is characterized by a set of 20 numbers (one
of three sets as explained above).
Then this set is multiplied by a single strength factor, depending
on the value of \evtspill.
As only additional complication, the strength factor has a $z$ and $R$
dependent ceiling to take into account the saturation.

The iterative procedure is terminated if the \dzeroprime curves of
 positive and negative pions are equal within $\pm$2~\mm over the whole
 spill. 
Typically, only one extra iteration is needed to obtain the required
 precision. 
This indicates that the characteristics of the \dzeroprime distributions
 describe the overall distortion strength reliably.
The result of the procedure is shown in Fig.~\ref{fig:be:dzeroprime}.
The small difference between the positive and negative pions around
 $\evtspill=50$ has no effect on the measurement of the momentum, but
 shows that the simple parabolic model describing the period of
 gradual onset of the distortions is not completely accurate.
The approximations used in the method are valid for values of 
$\langle \dzeroprime \rangle$ not exceeding 20~mm.
The shape of the $\langle \dzeroprime \rangle$ distribution as a
 function of \evtspill shows clearly up to which
value of event-in-spill the fitted parameters can be used. 
This maximum value is setting dependent, and is larger for beam settings
 which were better focused, for beams tuned at lower intensity and for
 targets of lower $z$.
The target material dependence is introduced by the multiplicity of the
 interaction products, which is higher for higher $z$.
In practice, this criterion does not represent a significant loss in final
 statistics of the data sets.
On average more the 80\% of the data can be reliably corrected.
The data sets which have had to be truncated most turn out to be the
 ones which were not statistics limited in any way
(e.g. the high $z$ data sets).

\section{Performance of TPC after correction}
\label{sec:corrected-performance}

In order to check the results of the corrections for the distortions
effects a number of control distributions were evaluated for each
analysed data-set.

\begin{figure}
\begin{center}
\includegraphics[width=0.43\textwidth]{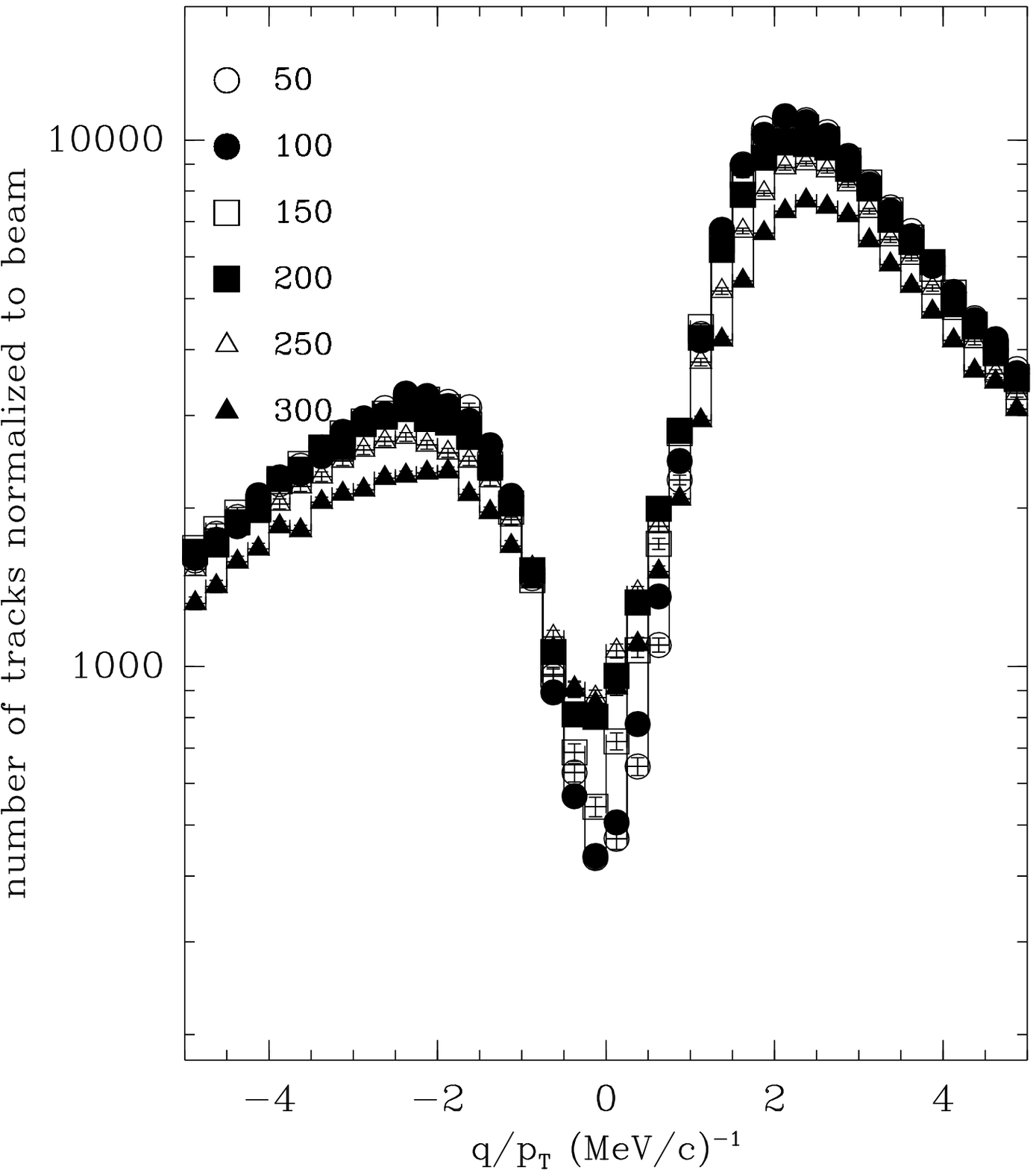}
~
\includegraphics[width=0.43\textwidth]{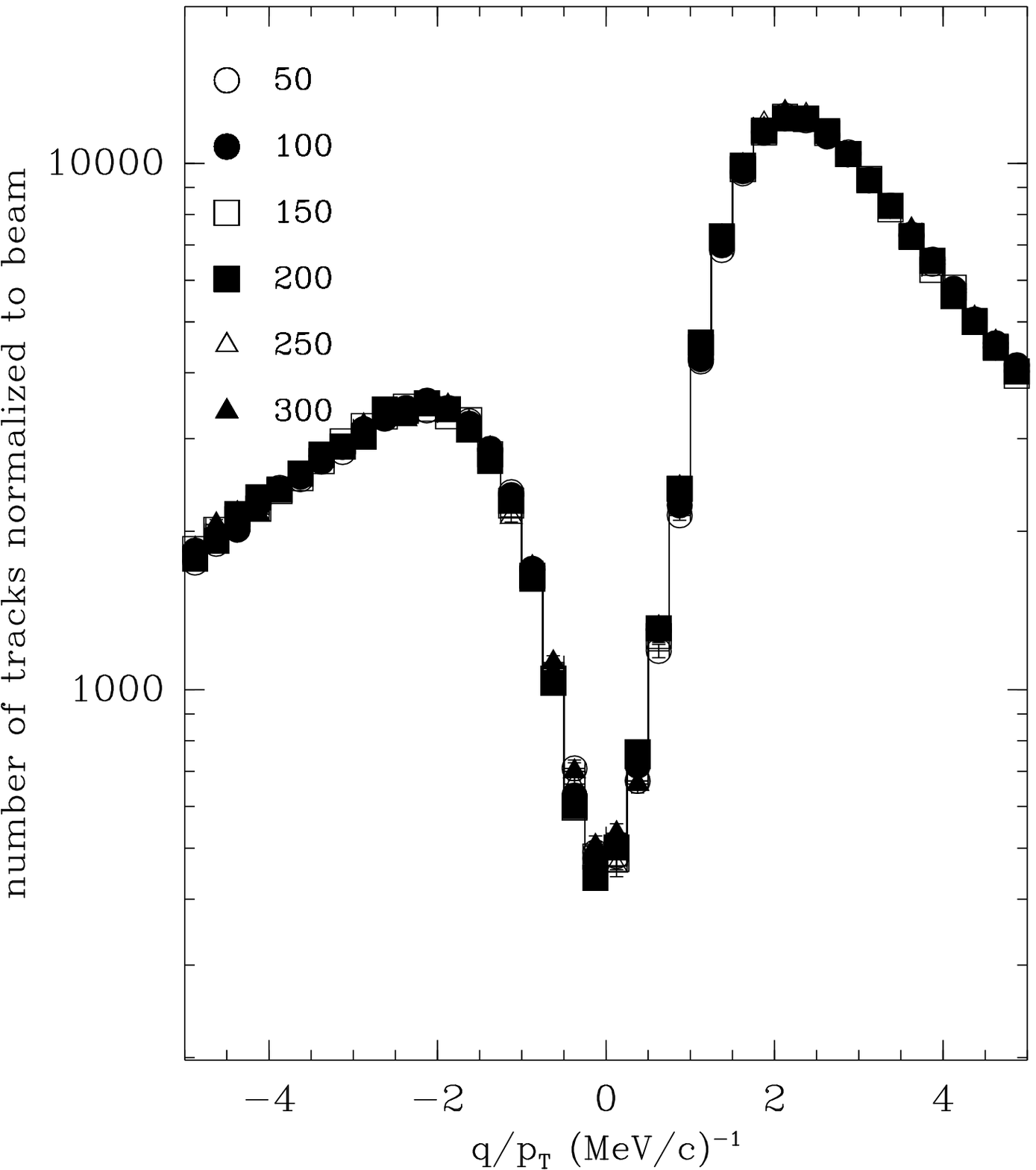}
\end{center}
\caption{
Distribution in $1/\pt$ for the 8.9~\GeVc Be data.  The six curves show
 six regions in event number in spill (each in groups of 50 events in
 spill). 
Groups are labelled with the last event number accepted in the group,
 e.g. ``50'' stands for the group with event number from 1 to 50.
The six groups are normalized to the same number of incoming beam
 particles, taking the first group as reference.
 Left panel: without dynamic distortion corrections; right panel: with
 dynamic distortion corrections.  
In the left panel only the first three groups of 50 events in spill are
 equivalent, while in the right panel all six groups are indistinguishable.
}
\label{fig:invpt}
\end{figure}

One control plot is the overall \pt distribution of all
tracks as a function of \evtspill.
Figure~\ref{fig:invpt} shows the distribution in $Q/\pt$, where $Q$ is
the measured charge of the particle, for six groups
of tracks, each corresponding $50 n < \evtspill \le 50 (n+1)$ (for $n$
ranging from zero to five).
The distributions have been normalized to an equal number of incident
beam particles, with the first group as reference.
In the left panel, no dynamic distortion correction have been applied
and a clear difference of the distributions is visible. 
One should note that the momentum measurement as well as the efficiency
is modified.  
The right panel shows the distributions after the corrections.
The distributions are no longer distinguishable.
To understand the asymmetry of positively and negatively charged tracks,
one should keep in mind that no particle identification was performed.
Thus both protons and pions contribute to the positives while the \pim's
are the only component of the negative particles.

\begin{figure}
\begin{center}
\includegraphics[width=0.45\textwidth,angle=0]{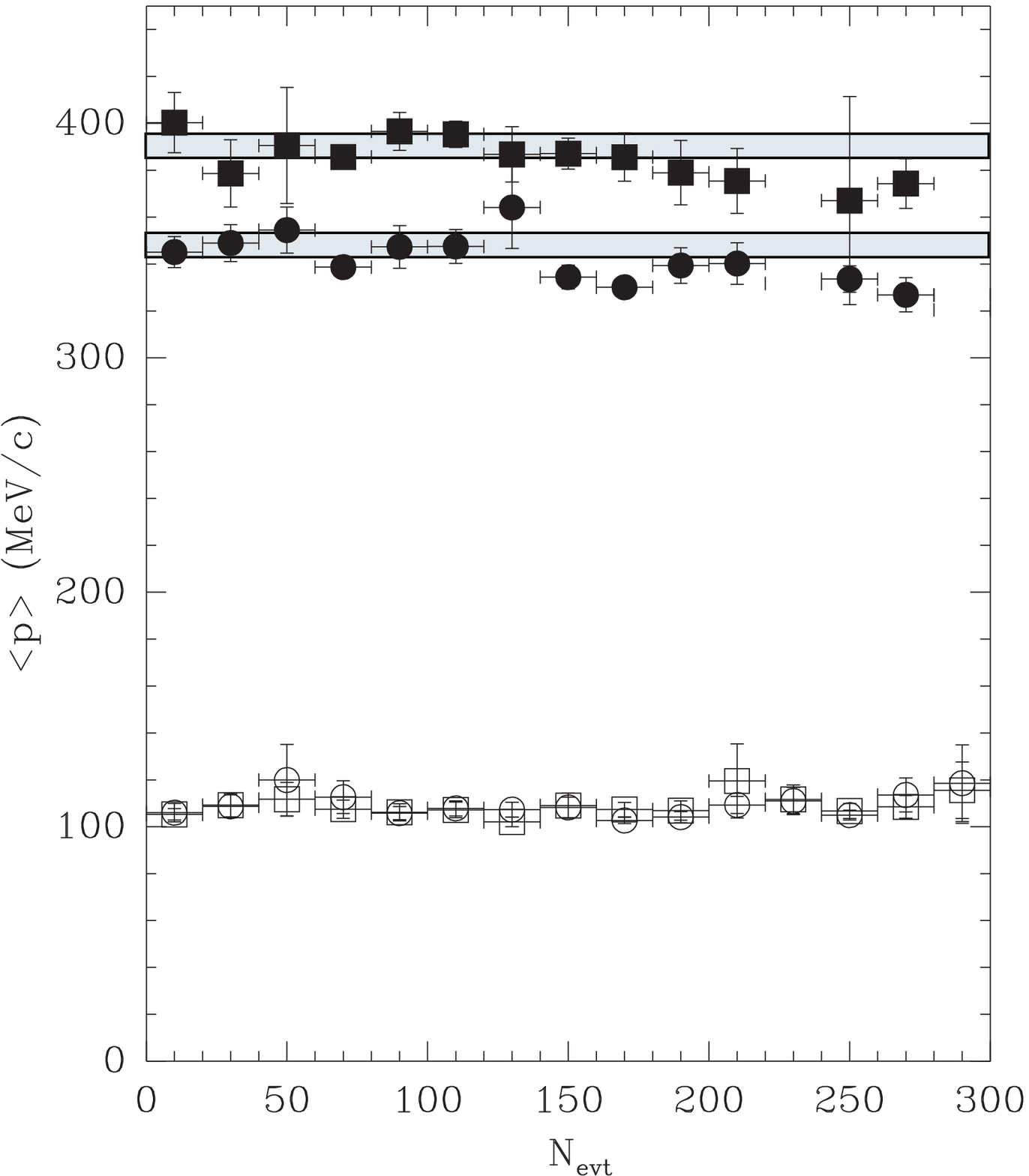}
~
\includegraphics[width=0.45\textwidth,angle=0]{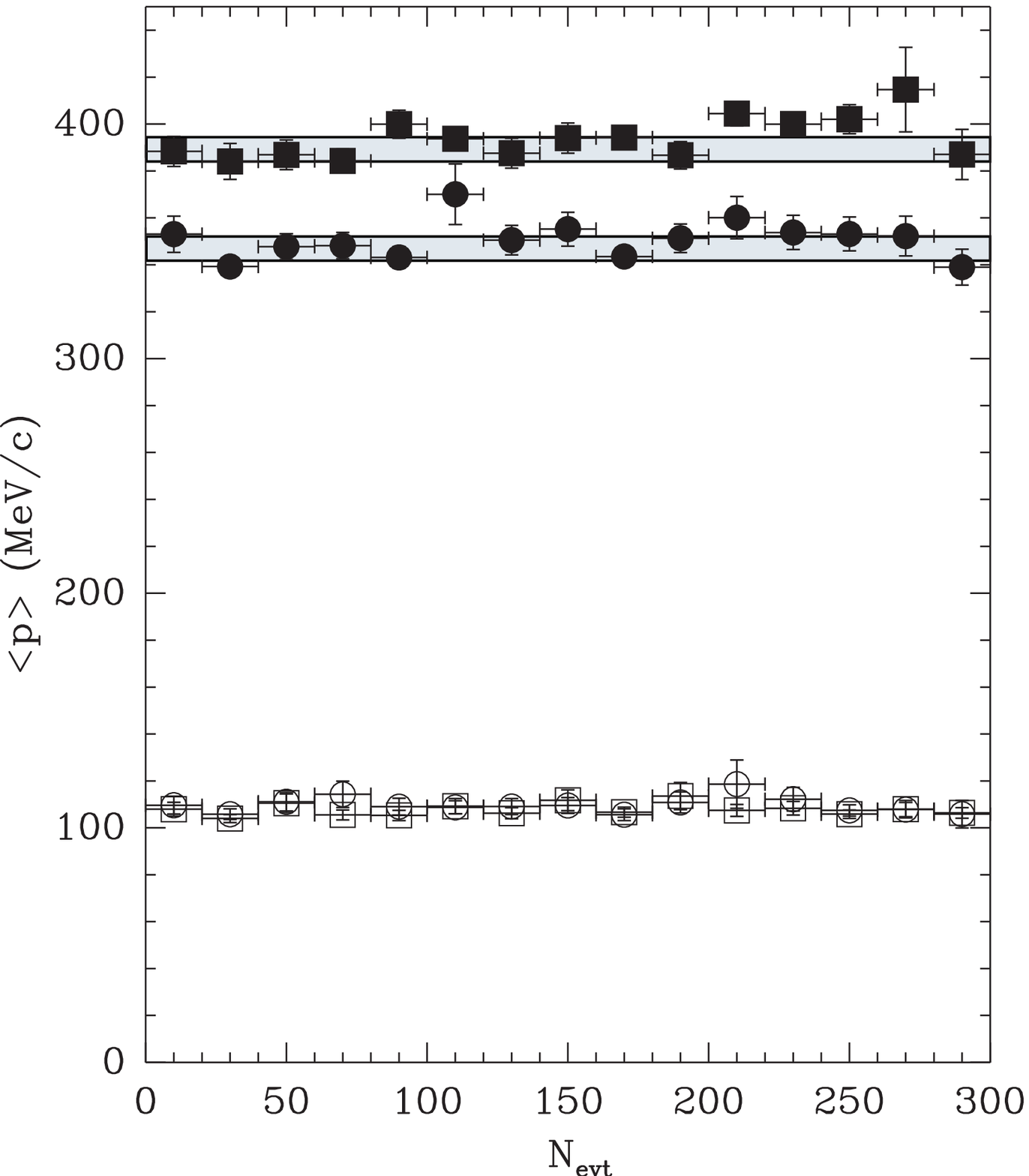}
\end{center}
\caption{
Momentum benchmarks. 
Left panel uncorrected; right panel: dynamic distortion
 corrections applied.
The closed boxes show the average momentum observed for protons selected
 using their range (reaching the second RPC) and \dedx; closed circles
 show protons selected within a high \dedx region; open circles: \pim
 selected with \dedx; open boxes: \pip selected with \dedx.
The angle of the particles is restricted in a range with $\sin \theta
 \approx 0.9$. 
In the left panel (uncorrected data) one observes a variation of
 $\approx 5\%$ for the high \pt samples.  
The corrected data stay stable well within 3\%. The shaded bands show a
 $\pm1.5$\% variation.
 The low \pt data remain stable with or without correction.
}
\label{fig:be:momentum}
\end{figure}

A more direct test of the effect of the  correction on the measurement
of momentum is shown in Fig.~\ref{fig:be:momentum}.
Four groups of tracks were selected, two classes of proton tracks and
pions separated on the basis of their charge.
A sample of relatively high momentum protons was selected
 using their range to set a lower limit.
The protons were required to produce a hit in two RPC layers.
A fixed window with relatively high values of \dedx in the TPC ensured
the particle identification as protons and limited the maximum momentum. 
Another window with higher values of \dedx selects protons with a lower
momentum. 
The pions are selected again by \dedx, which is only possible for low
momentum values (around 100~\MeV).
Positively charged and negatively charged pions are treated separately. 
The angle of the particles is restricted in a range with $\sin \theta
 \approx 0.9$, ensuring a small range of \pt. 
In the left panel (uncorrected data) one observes a variation of
 $\approx 5\%$ for the high \pt samples.  
The corrected data stay stable well within 3\%.
 The low \pt pion data remain stable with or without correction.
The width of the measured momentum distributions remains the same over
the length of the spill, indicating that also the resolution is well
corrected. 
It should also be noted that there is an effect on the efficiency.
While the efficiency to find a collection of clusters as a track is not
modified by the distortions, the requirement that the track is pointing
to the target does introduce an efficiency loss for the uncorrected
data. 
This loss is visible as an increase in the error bars on the
measurements.

From the combination of the two sets of control plots one can conclude
that the dynamic distortion corrections achieve a uniform efficiency and
a constant measurement of momentum over the whole spill.
Since the initial characterization of the TPC performance and
calibration was determined using the first part of the spill which is
not affected by dynamic distortions one expects that the calibration
remains applicable.
The systematic errors on these quantities remain approximately equal:
although an additional correction would improve the situation, more
events are now used with larger corrections applied to their tracks.

\section{Results for 8.9~\GeVc Be data}
\label{sec:cross-sections}

Finally, a comparison of the end-product of the analysis,
double-differential cross-sections, before and after the corrections can
be made.

\begin{figure}[tbp]
\begin{center}
\includegraphics[width=0.70\textwidth,angle=0]{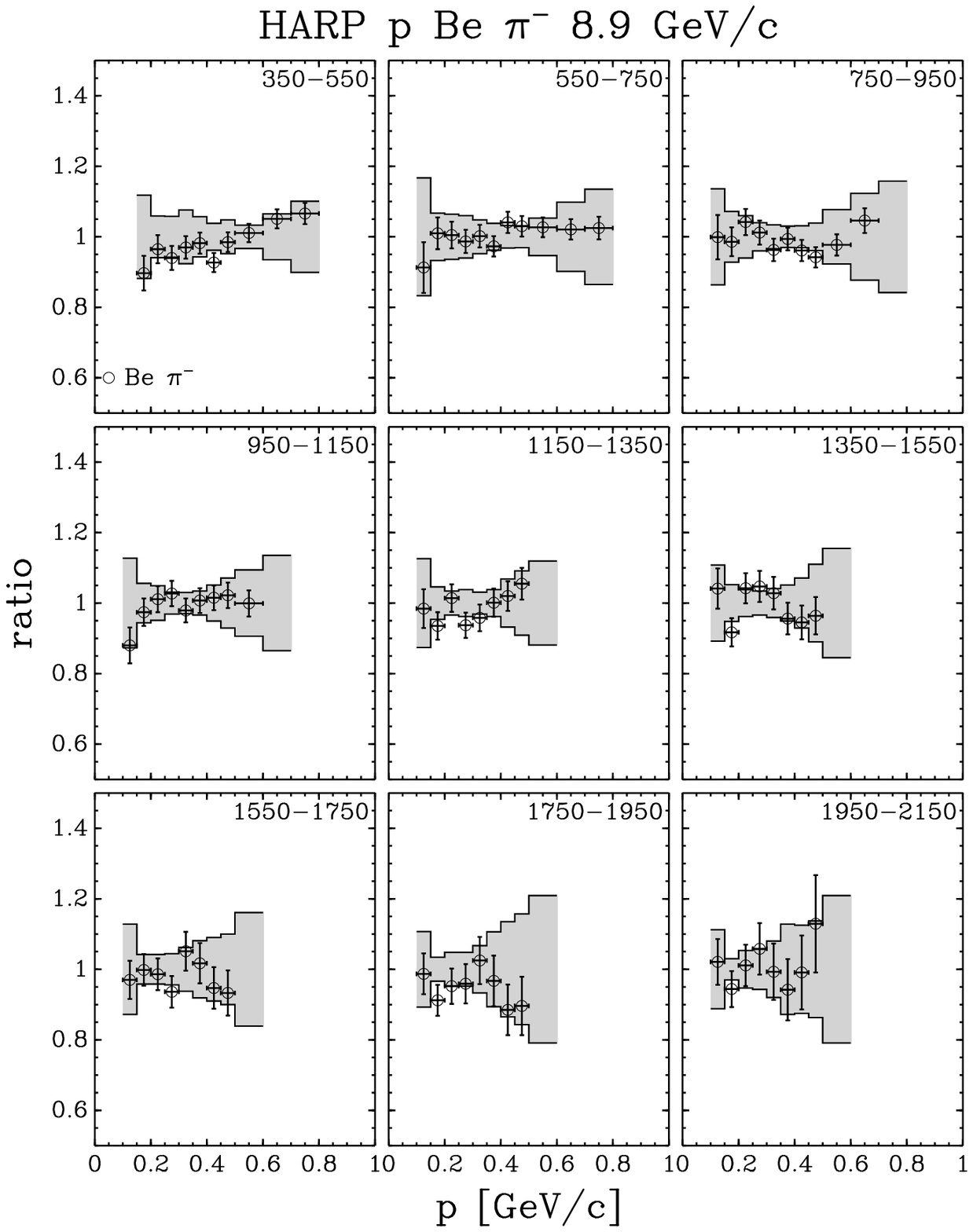}
\caption{Ratio of the \pim production cross-sections measured without and
with corrections for dynamic distortions in p--Be interactions at 8.9~\GeVc,
as a function of momentum shown in different angular bins (shown in mrad
in the panels). The error band in the ratio takes into account momentum
error and the error on the efficiency, the other errors being correlated.
The errors on the data points are statistical.
}
\label{fig:becomp1}
\end{center}
\end{figure}

\begin{figure}[tbp]
\begin{center}
\includegraphics[width=0.70\textwidth,angle=0]{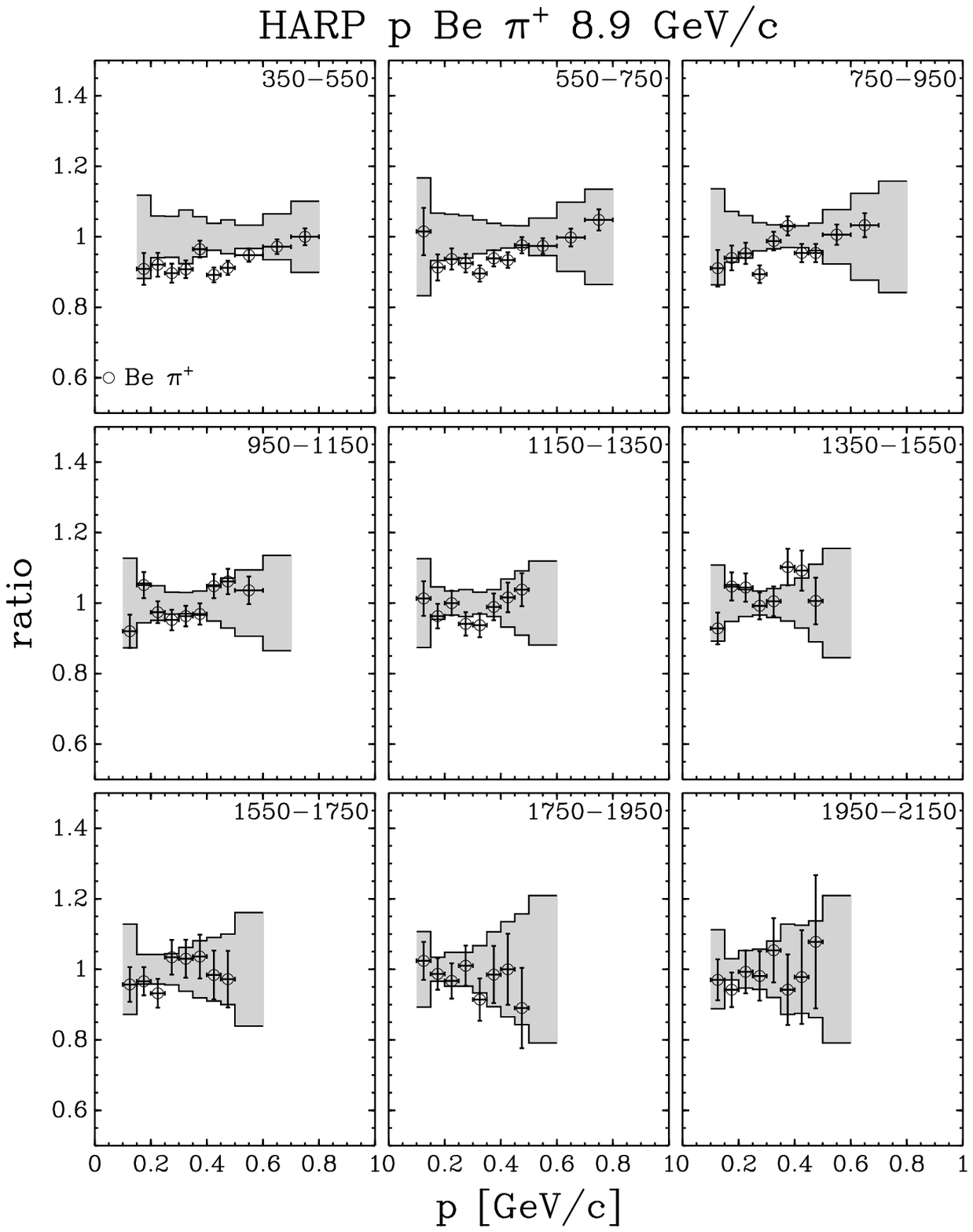}
\caption{Ratio of the \pip production cross-sections measured without and
with corrections for dynamic distortions in p--Be interactions at 8.9~\GeVc,
as a function of momentum shown in different angular bins (shown in mrad
in the panels). The error band in the ratio takes into account momentum
error and the error on the efficiency, the other errors being correlated.
The errors on the data points are statistical.
}
\label{fig:becomp2}
\end{center}
\end{figure}

The measured double-differential cross-sections for the 
production of \pip and \pim in the laboratory system as a function of
the momentum and the polar angle for each incident beam momentum were
measured for many targets and beam momenta.
These results are in agreement with what previously found using only
the first part of the spill and using no dynamic distortions 
corrections. 
Of course, both analyses only use the data for which their calibrations
are applicable.
Thus a lower statistics sample is used for the uncorrected data.
Making this comparison using the 8.9~\GeVc Be data has the advantage of
using the data set with the highest statistics, thus achieving the best
possible comparison.
Figures~\ref{fig:becomp1} and \ref{fig:becomp2} show the ratio of
the cross sections without and with the correction factor for dynamic
distortions in 8.9~\GeVc beryllium data. 
The error band in the ratio takes into account the usual estimate of
momentum error and the error on efficiency, the other errors are 
almost fully correlated. 
The agreement is within $1 \sigma$ for most of the points, confirming
the estimate of differential systematic error.
The statistical error bar represents the statistics of the
non-overlapping events.

\section{Conclusions}
\label{sec:conclusions}

The HARP TPC suffers a rather large number of operational problems.
The dynamic distortions observed for the particle trajectories
were tackled after the other problems had been corrected.  
The overall characteristics of the effect of these distortions were
described. 
Mainly the measurement of curvature and the extrapolation to the target
were affected.
It was shown that the origin of the distortions is fully understood both
theoretically and experimentally.
An experimental method to obtain a  direct measurement of the
distortions on the trajectory in space was developed.
The \dzeroprime variable has been identified to be a sensitive indicator of dynamic
distortions both with H$_2$ targets and heavier targets. 
The effect of dynamic distortions on the
particle  trajectories in the TPC has been measured
directly with H$_2$ targets by exploiting the forward
spectrometer and the kinematics of elastic scattering.
A simple model  of  the generation of  dynamic distortions 
and  a correction algorithm which depends on parameters that
are controlled by the \dzeroprime variable were developed. 
By monitoring the distortion strength with the \dzeroprime observable
the correction 
algorithm can be applied to all data sets taken with different targets.
The TPC performance (momentum scale and absolute efficiency) were
measured during the full spill by using data with hydrogen targets. 
The results of the corrections show that the performance of the TPC is
restored for the vast majority of the data.

\section{Acknowledgments}

We gratefully acknowledge the help and support of the PS beam staff
and of the numerous technical collaborators who contributed to the
detector design, construction, commissioning and operation.  
This work would not have been possible without the combined use of
all the detector components of HARP.
We would like to thank our colleagues from the HARP collaboration for
their contribution to this work.

 The experiment was made possible by grants from
the Institut Interuniversitaire des Sciences Nucl\'eair\-es, 
Ministerio de Educacion y Ciencia, Grant FPA2003-06921-c02-02 and
Generalitat Valenciana, grant GV00-054-1,
CERN (Geneva, Switzerland), 
the German Bundesministerium f\"ur Bildung und Forschung (Germany), 
the Istituto Na\-zio\-na\-le di Fisica Nucleare (Italy), 
INR RAS (Moscow), the Particle Physics and Astronomy Research Council
(UK) and the Swiss National Science Foundation, in the framework of the
SCOPES programme. 
We gratefully acknowledge their support.

\clearpage

\clearpage

\pagebreak

\pagebreak

\end{document}